\documentclass[showpacs,aps,amsmath,amssymb,twocolumn]{revtex4-2}

\usepackage[]{graphicx}
\usepackage{color}
\usepackage{bbold}
\usepackage{braket}
\usepackage[colorlinks=true, pdfstartview=FitV, linkcolor=blue, citecolor=red, urlcolor=black]{hyperref}
\newcommand{\be}{\begin{equation}}
\newcommand{\ee}{\end{equation}}
\newcommand{\ben}{\begin{eqnarray}}
\newcommand{\een}{\end{eqnarray}}
\newcommand{\bes}{\begin{subequations}}
\newcommand{\ees}{\end{subequations}}
\def\bal#1\eal{\begin{align}#1\end{align}}

\newcommand{\nn}{\nonumber\\}
\newcommand{\bfi}{\begin{figure}}
\newcommand{\efi}{\end{figure}}
\newcommand{\bc}{\begin{center}}
\newcommand{\ec}{\end{center}}
\newcommand{\sech}{\mbox{sech}}

\newcommand{\vphi}{\varphi}
\newcommand{\LL}{{\cal L}}
\newcommand{\Sc}{{\cal S}}
\newcommand{\Nc}{{\cal N}}

\begin{document}
\title{Generalized Jackiw-Teitelboim gravity in presence of Bloch brane-like models}
\author{I. Andrade$^{1}$, D. Bazeia$^{1}$, A. S. Lob\~ao Jr.$^{2}$, and R. Menezes$^{1,3}$}
\affiliation{$^1$Departamento de F\'\i sica, Universidade Federal da Para\'\i ba, 58051-970 Jo\~ao Pessoa, PB, Brazil\\
$^2$Escola T\'ecnica de Sa\'ude de Cajazeiras, Universidade Federal de Campina Grande, 58900-000 Cajazeiras, PB, Brazil\\
$^3$Departamento de Ci\^encias Exatas, Universidade Federal da Para\'\i ba, 58297-000 Rio Tinto, PB, Brazil}
\begin{abstract}
We investigate generalized Jackiw-Teitelboim gravity, coupling the dilaton field with two scalar matter fields. We obtain the equations of motion of the fields and investigate the linear perturbation of the solutions in general. We study two specific situations that allow analytic solutions with topological behavior and check how the dilaton field, the warp factor and Ricci scalar behave. In particular, we have shown how the parameters can be used to modify the structure of the solutions. Moreover, the perturbations are in general described by intricate coupled differential equations, but in some specific cases we could construct the corresponding zero modes analytically.
\end{abstract}
\maketitle

\section{Introduction}

An interesting way to study quantum effects in gravity is through two-dimensional gravitation. Although this is not a phenomenological description, $2D$ gravity allows testing some conjectures, providing a path for the construction of a quantum theory of gravity in the future. $2D$ gravity has also been used as a theoretical framework to describe complex situations of $4D$ gravity, such as evaporation of black holes \cite{Cruz:1996pg, Mertens:2019bvy, Cadoni:2021ypx}, dynamics of black holes \cite{Suh:2020lco}, supergravity \cite{Johnson:2020exp,Johnson:2020mwi, Fan:2021wsb} and other possibilities.

It is known that it is not possible to use the Einstein-Hilbert action for the description of $2D$ gravity, since it leads to identically null equations of motion. In this sense, it becomes necessary to resort to alternative representations to describe the two-dimensional gravity. A well known alternative was proposed in the 1980s by Jackiw  and Teitelboim, entitled Jackiw-Teitelboim (JT) gravity \cite{Teitelboim:1983,Jackiw:1985}. In this representation, a real scalar field coupled to gravity, called dilaton field, is used to provide the dynamics of the model. The dilaton field has been used to investigate other physical problems, not only through JT gravity, but also through other descriptions, see \cite{Alves:1999zna, Kummer:1999zy, Grumiller:2002nm}.

In order to encompass new possibilities, many proposals for generalizing the JT gravity have been presented over the years \cite{Mann:1991, Stotzel:1995, Grumiller:2021cwg, Mefford:2020vde, Momeni:2020zkx}. Many of them are motivated by the so-called modified theories of gravitation in $4D$, such as $F(R)$-gravity that introduce a general function of the Ricci scalar in the action \cite{Rippl:1995bg, Hwang:2001pu, DeFelice:2010aj, Sotiriou:2008rp}, Teleparallel Gravity, where curvature is replaced by torsion as the mechanism by which geometric deformation produces a gravitational field \cite{Bahamonde:2021gfp} and also $K$-fields, which includes modifications of the kinematics of the fields \cite{Armendariz-Picon:1999hyi, Armendariz-Picon:2000nqq}. Some generalized gravitation models have proved satisfactory in an attempt to build phenomenologically favorable inflationary models \cite{Starobinsky:1980, Steinhardt:1989, Olive:1990}.

Recently, new theoretical studies have deepened the discussion about the stability of topological solutions in generalized Jackiw-Teilelboim gravity. In Refs. \cite{Zhong:2021gxs, Zhong:2021hzu} it was shown that it is possible to investigate the linear stability of solutions by choosing an appropriate gauge. In Ref. \cite{Zhong:2021voa} it was shown that it is possible to obtain stable solutions for models with unusual dynamics in the form of $K$-fields, where cuscuton terms can be introduced to change stability conditions. More recently in Ref. \cite{Feng:2022ndn} the authors obtain double-kink solutions in models with standard dynamics.

The key point of these studies was the observation that the analysis of the stability is very similar to the scalar perturbations obtained in braneworld models in five-dimension, that are theories of gravity where the four-dimensional spacetime is immersed in an extra spatial dimension of infinite extent. This theory was proposed by Randall and Sundrum in 1999, and motivated to provide an alternative explanation for the hierarchy problem \cite{Randall:1999vf, Randall:1999ee}. The generalization of the original Randall-Sundrum scenario by incorporating scalar fields was initially proposed in \cite{Goldberger:1999uk, DeWolfe:1999cp, Csaki:2000fc}, and introduced new and interesting perspectives for brane cosmology, such as study of quintessence \cite{Huey:2001ae,Kunze:2001ji, Dias:2010}, inflation \cite{Hawkins:2000dq, Himemoto:2003jg, Himemoto:2002rq, Buchel:2004qg} and Teleparallel Gravity \cite{Menezes:2014bta, Geng:2014yya, Nozari:2012qi}; see \cite{Brito:2002xv, Bazeia:2015oqa, Bazeia:2015owa, Moreira:2021vcf, Moreira:2021xfe} for other extended braneworld scenarios.

As the $2D$ gravitation has been mirrored in the study of braneworld models, we think it is of interest to understand how the inclusion of new fields of matter can interfere in the study of stability. We know that, in braneworld scenario, when we include new scalar fields as source of density Lagrangian, there can appear interesting changes in the internal structure of the model. Furthermore, when we include the cuscuton term in Bloch brane models, it seems to induce the appearance of a split in warp factor \cite{Bazeia:2004dh,Bazeia:2021jok,Andrade:2018afh, Bazeia:2016uhr}.

Moreover, there is already an extensive literature studying topological defects in field theories in flat spacetime in the presence of several scalar fields \cite{Bazeia:1999xi, Bazeia:1998zv,Izquierdo:2002sz,Bazeia:2013uba,  Alonso-Izquierdo:2013isa}. We know that in these models the study of linear stability is not trivial, because the linearized field equations are, in general, coupled differential equations \cite{Bazeia:1997zp, Bazeia:1996}. Therefore, we think that replicating some considerations of the studies of field theories in flat spacetime, in this new scenario of $2D$ gravity, may also open new research issues for the dilaton gravity.

With these motivations on mind, in this work we organize the study  as follows. Sec. \ref{formaism} provide the general formalism that describe a generalized JT gravity in presence of dilaton field and also in presence of coupled scalar matter fields. In Sec. \ref{stability} we study the linear stability by using the dilaton gauge in the linearized equations of motion. In Sec. \ref{SpecificModels} we investigate two distinct models that engender kink-like solutions and describe conditions for the emergence of possible bound states. In Sec. \ref{coments} we present the conclusions and perspectives for future work.

\section{Formalism}\label{formaism}

Let us start this investigation considering a generalization of the Jackiw-Teitelboim gravity that describes a two-dimensional gravity in the form
\be\label{model}
\Sc= \frac{1}{\kappa}\!\int\!d^2x\sqrt{|g|}\left(\frac12\nabla_\mu\vphi\nabla^\mu\vphi-\vphi R+\kappa\,\LL_m\right),
\ee
where $\varphi$ is the dilaton field, $\kappa$ is a coupling constant, $g$ is the determinant of the metric $g_{\mu\nu}$, $R=g^{\mu\nu}R_{\mu\nu}$ is the Ricci scalar and $\LL_m$ is the Lagrangian density of matter. Here, the greek indexes $\mu,\nu,...$ run from $0$ to $1$ and the fields are all dimensionless.

It is possible to verify that the action defined by Eq.~\eqref{model} depends on several independents quantities which are, the dilaton field $\varphi$, the metric tensor $g_{\mu\nu}$ and in general the several fields introduced in the Lagrangian density of matter. In this sense, we can derive equations of motion for these quantities by variation of the action with respect to them. For example, by variation of action with respect to metric tensor we get the Einstein equation in the form
\be\label{Einst}
\begin{aligned}
\!\!g_{\mu\nu}\!\left(\nabla_{\!\alpha}\vphi\nabla^\alpha\vphi\!+\!4\Box \vphi\right)\!-\!2\nabla_{\!\mu}\vphi\nabla_{\!\nu}\vphi\!-\!4\nabla_{\!\mu}\!\nabla_{\!\nu}\vphi\!=\!2\kappa T_{\mu\nu},
\end{aligned}
\ee
where $\Box\equiv\nabla_\alpha\nabla^\alpha$ is the two-dimensional Laplacian operator and $T_{\mu\nu}$ is the energy-momentum tensor defined in the usual way as
\begin{equation}
    T_{\mu\nu}=\frac{2}{\sqrt{|g|}}\frac{\delta \left(\sqrt{|g|}\,\LL_m \right)}{\delta g^{\mu\nu}}.
\end{equation}
Note that, to obtain the specific form of energy-momentum tensor we must consider the Lagrangian density $\LL_m$. In this paper, we are interested in investigating models of two scalar fields as matter source fields. With that objective, we will consider a simple Lagrangian density that describes an interaction between the two fields $\psi$ and $\chi$ in the form
\be\label{Denslang}
\LL_m = \frac12\nabla_\mu\psi\nabla^\mu\psi+\frac12\nabla_\mu\chi\nabla^\mu\chi-V(\psi,\chi),
\ee
where $V(\psi,\chi)$ is the potential that govern the interaction of these fields. With this prescription, we can express the energy-momentum tensor as,
\be\nonumber
T_{\mu\nu}=\nabla_\mu\psi\nabla_\nu\psi+\nabla_\mu\chi\nabla_\nu\chi-g_{\mu\nu}\LL_m.
\ee
See that to close the representation giving by Lagrangian density \eqref{Denslang}  is also necessary to specific the form of potential. Making the variation of Eq.~\eqref{model} with respect to fields $\psi$ and $\chi$ and using the Lagrangian density \eqref{Denslang} we get,
\bes\label{eomS}
\ben
\nabla_\mu\nabla^\mu\psi+{V}_\psi = 0,\\
\nabla_\mu\nabla^\mu\chi+{V}_\chi = 0,
\een
\ees
where we use the indices in $V$ to denote derivative of potential with respect to the matter fields. Similarly, the equation of motion for the dilaton field is obtained by variation of Eq.~\eqref{model} with respect to $\vphi$, i.e.,
\ben\label{eomD}
\nabla_\mu\nabla^\mu\vphi+R=0.
\een
In this case, we then have four independent quantities to deal with: the dilaton, the metric tensor and the two scalar fields.

In an attempt to describe solutions with topological behavior, it was considered in \cite{Zhong:2021gxs} a two-dimensional representation of the Randall-Sundrum metric used to build five-dimensional braneworld models \cite{Randall:1999vf}. We will follow this line and consider a metric in the form
\be\label{metric}
ds^2=e^{2A}dt^2-dx^2.
\ee
As in brane models, $A$ is the warp function and $e^{2A}$ will also be called warp factor.  We assume that it depends only on the spatial coordinate $x$, i.e., $A=A(x)$. Thus, the Ricci scalar can be written as $R=2A'' +2{A^\prime}^2$, where the prime stands for the derivative with respect to $x$. Furthermore, we will consider static configurations for the matter fields and for the dilaton field, that is, $\psi=\psi(x)$, $\chi=\chi(x)$ and $\vphi=\vphi(x)$. In this case, the equations of motion \eqref{eomS} becomes,
\bes\label{seomS}
\bal
\psi'' +\psi^\prime A^\prime &= V_\psi,\label{seomSa}\\
\chi'' +\chi^\prime A^\prime &= V_\chi.\label{seomSb}
\eal
\ees
Note that, in general we get coupled equations for the fields $\psi$ and $\chi$ since $V$ depends on both fields. Using the static configurations, we can also obtain the non-vanishing components of Einstein equation \eqref{Einst} as
\bes\label{sEinst}
\bal
{\vphi^\prime}^2 +4\vphi'' &= -2\kappa\left(\frac12{\psi^\prime}^2 +\frac12{\chi^\prime}^2 +V\right),\label{sEinsta}\\
{\vphi^\prime}^2 -4A^\prime\vphi^\prime &= -2\kappa\left(\frac12{\psi^\prime}^2 +\frac12{\chi^\prime}^2 -V\right).\label{sEinstb}
\eal
\ees
On the other hand, the equation of motion for the dilaton field \eqref{eomD} becomes,
\be\label{seomD}
\vphi'' +\vphi^\prime A^\prime = 2A'' +2{A^\prime}^2.
\ee

The five differential equations represented by Eqs.~\eqref{seomS}, Eqs.~\eqref{sEinst} and Eq.~\eqref{seomD} describe all known information about the system. It is possible to show that one of these equations is not independent and can be obtained from the others, for example, we can use Eqs. \eqref{seomS}, \eqref{sEinstb} and \eqref{seomD} to obtain \eqref{sEinsta}. Thus, the set of five equations reduces to four independents equations. This is all we need, since here we have the dilaton $\varphi$, the warp function $A$ and the two scalars $\psi$ and $\chi$ to be determined.

It was shown in \cite{Zhong:2021gxs} that it is possible obtain a general solution for Eq. \eqref{seomD} in terms of two integration constants. In order to deal with first-order equations we consider in this paper a particular solution of Eq. \eqref{seomD} in the form,
\be\label{solD}
\vphi(x) = 2A(x).
\ee
To improve the mathematical description, we can use the above solution to rewrite the Eqs.~\eqref{sEinst} as
\be\label{Einst2o}
-4A'' = \kappa{\psi^\prime}^2 +\kappa{\chi^\prime}^2,
\ee
and
\be\label{potx}
4{A^\prime}^2 = \kappa{\psi^\prime}^2 +\kappa{\chi^\prime}^2 -2\kappa V.
\ee
From these two equations it is possible to write the Ricci scalar defined below Eq.~\eqref{metric} in terms of the potential as 
\be\label{Ricci}
R=-\kappa V.
\ee
Note that now we need to work with a system of second-order differential equations, that can be solved using the so-called first-order formalism allowing to reduce second-order differential equations to first-order equations. To proceed with this method, we must introduce an auxiliary function $W(\psi,\chi)$ that correlates the fields $\psi$ and $\chi$ such that,
\be\label{fo}
\psi^\prime = W_\psi \quad\text{and}\quad \chi^\prime = W_\chi,
\ee
where $W_{\psi}=\partial W/\partial \psi$ and $W_{\chi}=\partial W/\partial \chi$. Using the prescription in Eq. \eqref{Einst2o} we get the warp function as
\be\label{foA}
A^\prime = -\,\frac{\kappa}{4}\,W(\psi,\chi).
\ee
Moreover, we can use Eq. \eqref{potx} to write the potential in the form
\begin{equation}\label{potW}
    V(\psi,\chi)=\frac12 \,W_{\psi}^2+\frac12\, W_{\chi}^2-\frac{\kappa}{8}\, W^2.
\end{equation}

The set of first-order equations represented in \eqref{fo} are commonly found when one studies models described by two scalar fields. See for example \cite{Bazeia:1999xi,Alonso-Izquierdo:2013isa} where the authors studied the presence of kink-like solutions in two-dimensional Minkowski spacetime. Or also in \cite{Bazeia:2004dh}, where it was investigated a system described by two real scalar fields coupled with gravity in $(4, 1)$ dimensions in warped spacetime involving one extra dimension. It is worth noting that the first-order equations \eqref{fo} and \eqref{foA} solve the Eqs. \eqref{seomS} and \eqref{sEinst} provided that the potential is given by \eqref{potW}. Furthermore, static and uniform solutions can be obtained from the algebraic equations $W_\psi=0$ and $W_{\chi}=0$, which takes us to a set of points in the space of fields given by $v_i=(\bar\psi_i,\bar\chi_i)$, $i=1,2,\cdots$, that satisfies the Eqs. \eqref{fo}. Then, we impose that the static solutions $\psi(x)$ and $\chi(x)$ tend to these values when $x\to\pm\infty$.

Due to the asymptotic behavior of the fields, as described above, the function $W$ assumes constant values when the fields $\psi(x)$ and $\chi(x)$ are evaluated at $x\to\pm \infty$, i.e., $W(\psi(x\to\pm\infty),\chi(x\to\pm\infty))=W_{\pm}$. With this, we can use the Eq. \eqref{foA} to write $A(|x|\!\gg\!0)\approx -(\kappa W_\pm/4)\,x$, such that the warp factor $e^{2A}$ becomes, asymptotically,
\be\label{Aasympt}
e^{-(\kappa W_\pm/2)\,x}.
\ee
Thus, it may diverge, become a positive constant or vanish, depending on the sign of $W_\pm$ for $\kappa$ positive.

\section{Linear Stability}\label{stability}

In this section we study the linear stability of dilaton gravitation in presence of matter fields considering small perturbations around static solutions of fields. Firstly, let us consider small perturbations in matter fields in the form $\psi\to\psi(x)+\eta(x,t)$ and $\chi\to\chi(x)+\xi(x,t)$. For the dilaton field we write $\vphi\to \vphi(x)+\delta \vphi(x,t)$. Lastly, we consider perturbations in metric tensor as $g_{\mu\nu}\to g_{\mu\nu}(x)+\pi_{\mu\nu}(x,t)$, where the indices of $\pi_{\mu\nu}$ are raised or lowered as $\pi^{\mu\nu}= -g^{\mu\alpha}\pi_{\alpha\beta}g^{\beta\nu}$.

Using the field perturbations, we can linearize the equations of motion to investigate the linear stability. For example, the $(0,0)$ component of Einstein equation \eqref{Einst} can be written as
\be
\begin{aligned}\label{comp00}
\!\!\!\!\!&-2\vphi^\prime\delta\vphi^\prime -4\delta\vphi'' -{\vphi^\prime}^2\pi_{11} -4\vphi''\pi_{11} -2\vphi^\prime\pi_{11}^\prime\\
\!\!\!\!\!&= 2\kappa\!\left(\!V_\psi\eta \!+\!V_\chi\xi \!+\!\psi^\prime\eta^\prime \!+\!\chi^\prime\xi^\prime \!+\!\frac12{\psi^\prime}^2\pi_{11} \!+\!\frac12{\chi^\prime}^2\pi_{11}\!\right)\!,
\end{aligned}
\ee
where $V_\psi$ and $V_\chi$ are applied to the static solutions. The $(0,1)$ or $(1,0)$ components are identical and have the form
\be
2A^\prime\delta\vphi -\vphi^\prime\delta\vphi -2\delta\vphi^\prime -\vphi^\prime\pi_{11} = \kappa\left(\psi^\prime\eta +\chi^\prime\xi\right).
\ee
On the other hand, the $(1,1)$ component is
\be
\begin{aligned}\label{comp11}
&\!\!\!\!4e^{-2A}\ddot{\delta\vphi} \!-\!4A^\prime\delta\vphi^\prime \!+\!2\vphi^\prime\delta\vphi^\prime \!+\!4\vphi^\prime\Pi \!-\!{\vphi^\prime}^2\pi_{11} \!-\!4\vphi''\pi_{11}\\
&\!\!\!\!=\! 2\kappa \left(\!V_\psi\eta \!+\!V_\chi\xi \!-\!\psi^\prime\eta^\prime \!-\!\chi^\prime\xi^\prime \!+\!\frac12{\psi^\prime}^2\pi_{11} \!+\!\frac12{\chi^\prime}^2\pi_{11}\!\right),
\end{aligned}
\ee
where we used the dot to express the derivative with respect to $t$ and introduced a new variable as
\be
\Pi = e^{-2A}\left(\dot{\pi}_{01} +A^\prime\pi_{00} -\frac12\pi_{00}^\prime\right).
\ee

One can show that the linearization of the equations of motion \eqref{eomS} provides us with the relationships,
\be\label{stabSa}
\begin{aligned}
   &e^{-2A}\ddot{\eta} -e^{-A}\left(e^A\eta^\prime\right)^\prime +V_{\psi\psi}\eta +V_{\psi\chi}\xi\\
   &+\psi^\prime\Pi -A^\prime\psi^\prime\pi_{11} -\psi''\pi_{11} -\frac12\psi^\prime\pi_{11}^\prime = 0,
\end{aligned}
\ee
and
\be\label{stabSb}
\begin{aligned}
    &e^{-2A}\ddot{\xi} -e^{-A}\left(e^A\xi^\prime\right)^\prime +V_{\chi\psi}\eta +V_{\chi\chi}\xi\\
    &+\chi^\prime\Pi -A^\prime\chi^\prime\pi_{11} -\chi''\pi_{11} -\frac12\chi^\prime\pi_{11}^\prime = 0.
\end{aligned}
\ee

We can also make the linearization of the dilaton equation \eqref{eomD}. Here, however, we will follow the prescription used in \cite{Zhong:2021gxs} and adopt the dilaton gauge, i.e., $\delta \vphi=0$. Whit this choice, the linearized equation that comes from Eq. \eqref{eomD} vanish. Furthermore, we can use the Eq. \eqref{solD} to write the Eqs. \eqref{comp00} and \eqref{comp11}, respectively, as
\bes\label{vinc}
\bal
\pi_{11} &= -\frac{\kappa}{2A^\prime}\big(\psi^\prime\eta +\chi^\prime\xi\big),\\
\Pi &= \frac{\kappa}{4}\left(\left(\frac{\psi^\prime}{A^\prime}\right)^\prime\!\eta \!+\!\left(\frac{\chi^\prime}{A^\prime}\right)^\prime\!\xi \!-\!\frac{\psi^\prime}{A^\prime}\eta^\prime \!-\!\frac{\chi^\prime}{A^\prime}\xi^\prime\right).
\eal
\ees

Let us assume that the perturbations in matter fields can be decomposed as $\eta(x,t)=\sum_n\eta_n(x)\cos(\omega_nt)$ and $\xi(x,t)=\sum_n\xi_n(x)\cos(\omega_nt)$, where $\omega_n$ is a characteristic frequency. Using this decomposition and the set of Eqs. \eqref{vinc}, we can represent the Eqs.~\eqref{stabSa} and \eqref{stabSb} as
\be\label{stabSL}
-e^A\left(e^A\Upsilon^\prime_n\right)^\prime +e^{2A}U(x)\Upsilon_n = \omega_n^2\Upsilon_n\,,
\ee
where we defined
\be
U(x) =
\begin{pmatrix}
p(x) & q(x) \\
q(x) & \bar{p}(x)
\end{pmatrix}
 ,
\quad\quad
\Upsilon_n =
\begin{pmatrix}
\eta_n(x) \\
\xi_n(x)
\end{pmatrix},
\ee
and
\bes\label{compstabSL}
\bal
p(x) &= V_{\psi\psi} +\frac{\kappa}{2}\left({\psi^\prime}^2 +\left(\frac{{\psi^\prime}^2}{A^\prime}\right)^{\!\prime}\,\right),\\
\bar{p}(x) &= V_{\chi\chi} +\frac{\kappa}{2}\left({\chi^\prime}^2 +\left(\frac{{\chi^\prime}^2}{A^\prime}\right)^{\!\prime}\,\right),\\
q(x) &= V_{\psi\chi} +\frac{\kappa}{2}\left(\psi^\prime\chi^\prime +\left(\frac{\psi^\prime\chi^\prime}{A^\prime}\right)^{\!\prime}\,\right).
\eal
\ees
Note that the Eq. \eqref{stabSL} is a Sturm-Liouville equation. We can define the inner product of two states as
\be\label{norma}
\braket{\Phi|\Upsilon} = \int dx\,\rho(x)\Phi^\dagger(x)\Upsilon(x),
\ee
where $\rho(x)=e^{-A(x)}$ is the weight function \cite{Hounkonnou:2004,andrade:2020}. One can show that the Eq. \eqref{stabSL} has a state with $\omega=0$, which is given by
\be\label{zeroSL}
\Upsilon^{(0)}(x) = \frac{\Nc}{A^\prime}
\begin{pmatrix}
\psi^\prime \\
\chi^\prime \\
\end{pmatrix},
\ee
where $\Nc$ is a normalization constant which can be determined from the equation \eqref{norma}.

We can use the first-order equations \eqref{fo}, \eqref{foA} and the potential in the form \eqref{potW} to rewrite the Eq. \eqref{stabSL} in terms of the function $W(\psi,\chi)$ as
\be\label{estabilytyM}
\!\!-e^A\!\left(e^A\Upsilon^\prime_n\right)^\prime +e^{A}\!\left(e^AM^2 \!+\!\left(e^AM\right)^\prime\right)\!\Upsilon_n = \omega_n^2\Upsilon_n\,,
\ee
where we defined
\be
M =
\begin{pmatrix}
W_{\psi\psi} -\frac{W_\psi^2}{W} & W_{\psi\chi} -\frac{W_\psi W_\chi}{W} \\
W_{\chi\psi} -\frac{W_\chi W_\psi}{W} & W_{\chi\chi} -\frac{W_\chi^2}{W} \\
\end{pmatrix}.
\ee
In this case, we can express the stability equation as $S^\dagger S\Upsilon_n=\omega_n^2\Upsilon_n$, where
\be\label{Sop}
\!S = e^A\!\left(\!-\frac{d}{dx}\mathbb{1} +M\!\right); \quad S^\dagger = e^A\!\left(\!\frac{d}{dx}\mathbb{1} +M\!\right)\!.
\ee
As in the study of the supersymmetric quantum mechanics \cite{cooper:1995}, we can define the supersymmetric partner operator as $SS^\dagger$, and applying in the state $\Phi_n$ we will have
\be\label{partnersusy}
SS^\dagger\Phi_n \!=\! -e^A\!\left(e^A\Phi^\prime_n\right)^\prime \!+e^{A}\!\left(e^A\!M^2 -\left(e^A\!M\right)^\prime\right)\!\Phi_n.
\ee
The supersymmetric partner operators $S^\dagger S$ and $SS^\dagger$ can be used to relate their respective eigenstates and eigenvalues, which can facilitate the study of the stability equation \eqref{stabSL}.

We can also make a change of variable in the form $dz=e^{-A}dx$, in order to make the metric conformally flat; in this case, the stability equation \eqref{stabSL} becomes a Schr\"odinger-like equation, i.e.,
\be\label{esSchr}
-\frac{d^2\Upsilon_n}{dz^2} +{\cal U}(z)\Upsilon_n = \omega_n^2\Upsilon_n,
\ee
where
\be
{\cal U}(z) \!=\!\!
\begin{pmatrix}
    e^{2A}V_{\psi\psi}\! +\!\frac{\kappa}{2}\!\left(\frac{\psi_z^2}{A_z}\right)_{\!z} & e^{2A}V_{\psi\chi} \!+\!\frac{\kappa}{2}\!\left(\!\frac{\psi_z\chi_z}{A_z}\!\right)_{\!z} \\
    e^{2A}V_{\chi\psi} \!+\!\frac{\kappa}{2}\!\left(\!\frac{\chi_z\psi_z}{A_z}\!\right)_{\!z} & e^{2A}V_{\chi\chi} \!+\!\frac{\kappa}{2}\!\left(\frac{\chi_z^2}{A_z}\right)_{\!z}\label{PotStaMatrix}
\end{pmatrix}\!\!.
\ee
Here, we are using the index $z$ to represent derivative with respect to the new variable $z$, as in $\psi_z=d\psi/dz$, etc.  We also have a state with $\omega=0$, that is
\be\label{modzerosho}
\Upsilon^{(0)}(z) = \frac{\Nc}{A_z}
\begin{pmatrix}
    \psi_z \\
    \chi_z \\
\end{pmatrix}.
\ee
Similarly to the Sturm-Liouville equation, we can fatorize the Eq.~\eqref{esSchr} as $\Sc^\dagger\Sc\,\Upsilon_n=\omega_n^2\Upsilon_n$. Here the operator $\Sc$ is given by
\be
\!\Sc \!=\!\!
\begin{pmatrix}
    \!-\frac{d}{dz} \!+\! e^A\!\left(\!W_{\psi\psi} \!-\!\frac{W_\psi^2}{W}\!\right) & e^A\!\left(\!W_{\psi\chi} \!-\!\frac{W_\psi W_\chi}{W}\right) \\
    e^A\!\left(\!W_{\chi\psi} \!-\!\frac{W_\chi W_\psi}{W}\!\right) & \!-\frac{d}{dz} \!+\! e^A\!\left(\!W_{\chi\chi} \!-\!\frac{W_\chi^2}{W}\!\right) \\
\end{pmatrix}\!\!.
\ee

As we see, it is possible to study the linear stability of static solutions through a Schr\"odinger-like equation. In this case, to write the correspondence between the two variables $x$ and $z$, one has to integrate to find $x$ as a function of $z$. However, this change cannot always be done analytically. Therefore, it is necessary to resort to numerical methods, as we will illustrate in one of our examples.

We can see from Eq. \eqref{modzerosho} that the zero mode may be divergent for $A_z=0$, and this may lead to a non normalized zero mode. Since the derivative of the warp function is proportional to $W$ (see Eq. \eqref{foA}), we can analyze the asymptotic behavior of $W$ to get further insight on the behavior of the zero mode. We know that $W\to W_{\pm}$ asymptotically, so if the sign of $W_-$ is different from the sign of $W_+$, $W$ has to vanish somewhere in the $z$ axis to obstruct the normalization of the zero mode. In this sense, to make the zero mode normalizable, the sign of $W$ should not change, and the warp factor should not diverge asymptotically.

\section{Specific models}\label{SpecificModels}

In this section, we study two distinct models in order to understand how the formalism presented so far works. Furthermore, we investigate how the matter fields interact and give rise to the dilaton field, the warp factor and the Ricci scalar.

\subsection{Model A}\label{modA}

The first model that we consider in this paper is motivate by the so-called Bloch brane that is five-dimension braneworld model constructed by interaction of two real scalar fields. The model was firstly considered in \cite{Bazeia:2004dh}, where the authors used an auxiliary function $W(\psi,\chi)$, in which we now add a real constant $c$ to get
\be\label{WBNRT}
W(\psi,\chi) = c +\psi -\frac{\psi^3}{3} -r\psi\chi^2,
\ee
where $r$ is real parameter. In particular, $c$ can be used to modify the structure of the solutions and $r$ is positive and controls the coupling between the fields. Using the algebraic equations, $W_\chi=0$ and $W_\psi=0$, we obtain four sets of values to the asymptotic behavior of fields: $v_{1}=(1,0)$, $v_{2}=(-1,0)$, $v_{3}=(0,1/\sqrt{r})$ and $v_{4}=(0,-1/\sqrt{r})$. Therefore, the auxiliary function $W$ assume the values, $W(v_{1})\!=\!c+2/3$, $W(v_{2})\!=\!c-2/3$ and $W(v_{3})\!=\!W(v_{4})\!=\!c$.

See that we can obtain the asymptotic behavior of warp factor using the asymptotic behavior of fields. For example,  for the solutions that connect $v_2$ to $v_1$, we get
\be
e^{2A\left(|x|\gg0\right)} \approx e^{-(\kappa/2)(2/3\pm c)|x|},
\ee
where the plus sign in the exponential is for $x\to\infty$ and the minus sign for  $x\!\to\!-\infty$. Note that, the warp factor has an asymmetric behavior when $c\!\neq\! 0$; furthermore, it tends asymptotically to zero at both extremes if $|c|\!<\!2/3$, however, if $|c|\!>\!2/3$ the warp factor diverges at one side. On the other hand, the solutions that connect the values $v_4$ and $v_3$, leads to
\be
e^{2A\left(|x|\gg0\right)} \approx e^{-\kappa cx/2}.
\ee
In this case, the warp factor is asymmetric and not localized anymore. To the model defined in \eqref{WBNRT}, we can obtain the interaction potential of the matter fields as
\be\label{VBNRT}
\begin{aligned}
V(\psi,\chi) =\,& \frac{1}{2}\left(1-\psi ^2-r \chi ^2\right)^2 +2r^2\psi^2 \chi^2\\
	&-\frac{\kappa}{8}\left(c+\psi-\frac{\psi^3}{3}-r\psi  \chi^2\right)^2.
\end{aligned}
\ee
Using the asymptotic values of the solutions we find that $V(\pm1,0)\! =\! -(\kappa/8)(c\!\pm\!2/3)^2$ and $V(0,\pm 1/\sqrt{r}) \!=\! -\kappa c^2/8$. Note that, for $\kappa>0$ we have $V(\bar{\psi}_i,\bar{\chi}_i)\leq0$ in all cases; this indicates that the space can be $AdS_2$ or $M_2$ asymptotically depending on the value of parameter $c$. As mentioned below Eq.~\eqref{potx}, it is possible to relate the potential with the Ricci scalar. This result is interesting because it is possible to calculate the asymptotic value of the Ricci scalar directly without knowing the solutions in their explicit form, using only the required boundary conditions, as we will implement below.

We can now investigate the specific solutions of model. For this, we use the Eqs. \eqref{fo} to obtain a system of first-order differential equations as
\bes\label{eqMAacopladas}
\bal
\psi^\prime &=1-\psi^{2}-r\chi^{2},\\
\chi^\prime &=-2r\psi\chi.
\eal
\ees

\begin{figure}[!htb]
    \begin{center}
        \includegraphics[scale=0.6]{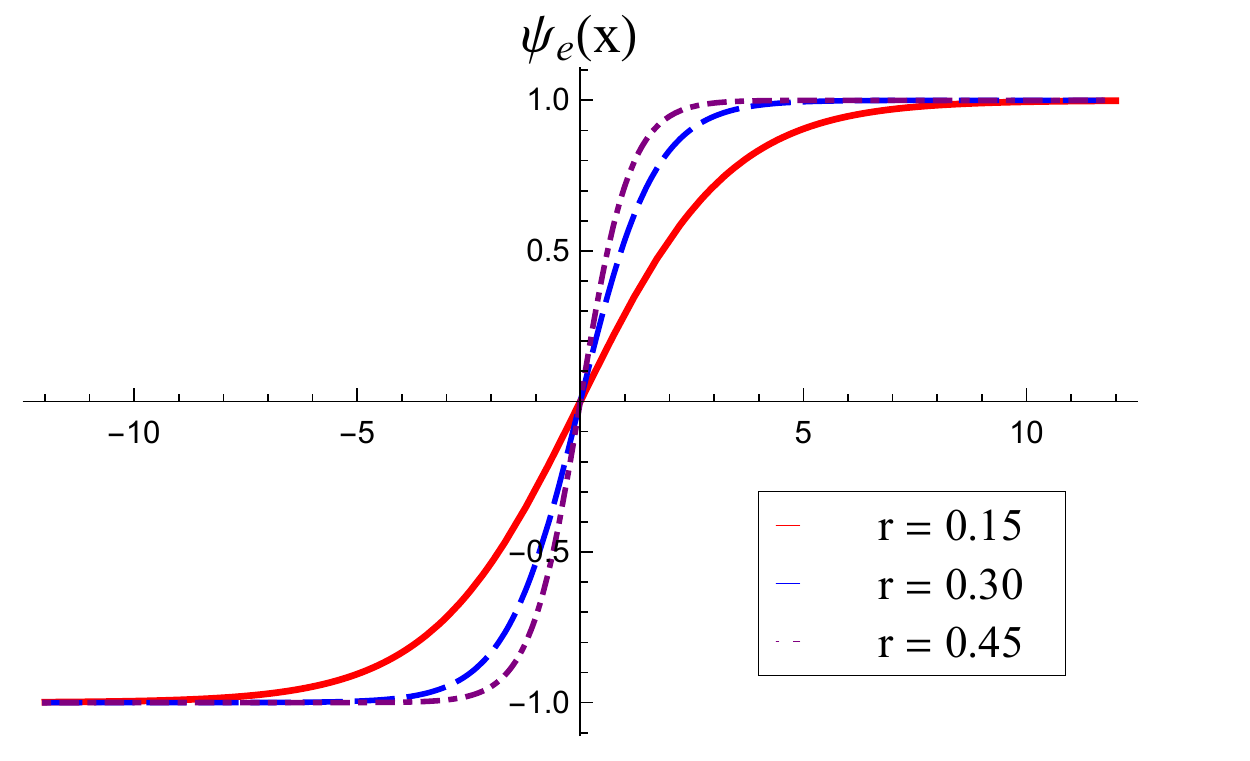}
        \includegraphics[scale=0.6]{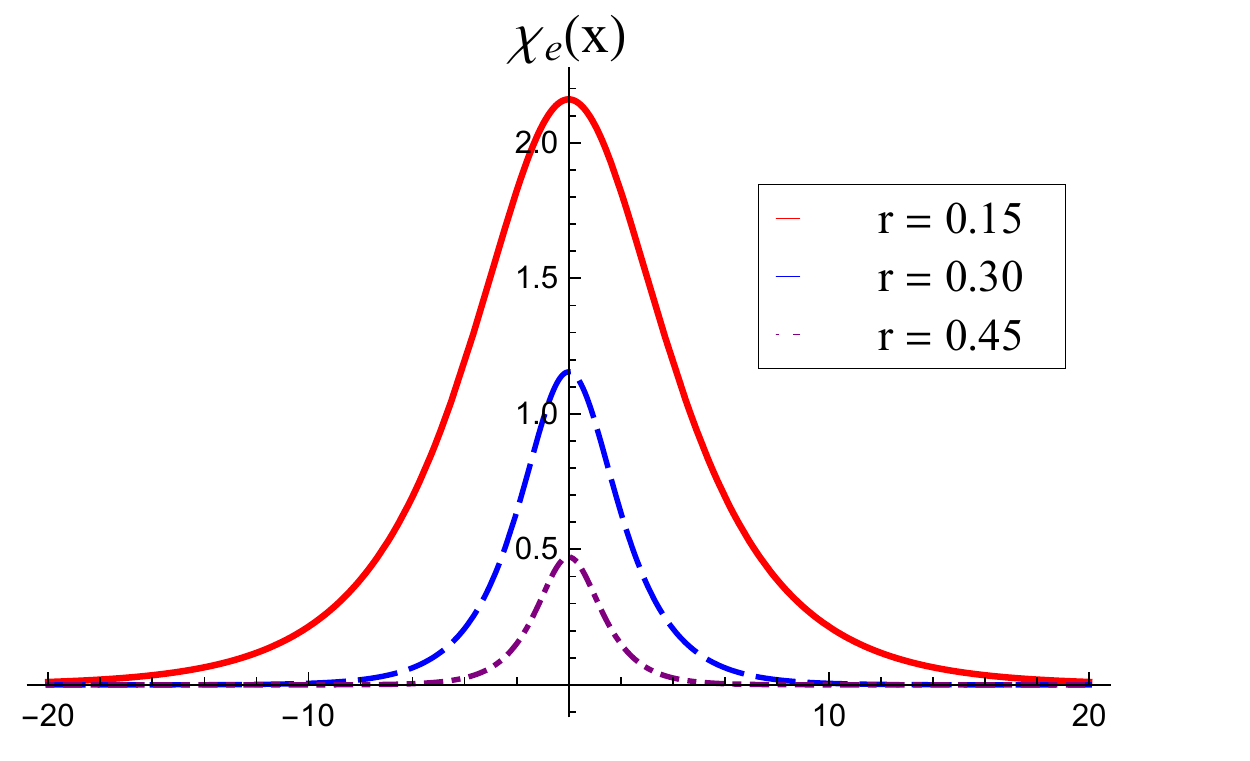}
    \end{center}
    \vspace{-0.5cm}
    \caption{\small{Solutions of the elliptic orbit represent by Eq. \eqref{solOelipA} (top panel) and Eq. \eqref{solOelipB} (bottom panel).}\label{fig01}}
\end{figure}

We can solve numerically the set of coupled differential equations \eqref{eqMAacopladas} aiming to obtain solutions that connect the values $v_i$ obtained by solving the algebraic equations. However, it was shown in \cite{Izquierdo:2002sz} that it is possible to decouple this system of equations considering orbits $F(\psi,\chi)=0$ that connect the uniform solutions $v_i$. For this model we have the orbits in the form
\be\label{Oelliptic}
\psi^{2}+\left(\frac{r}{1\!-\!2r}\right)\chi^{2}-b\,\chi^{1/r}=1,
\ee
where $b$ is a real integration constant that controls the shape of the above orbits. Here we consider $b=0$. There is an interesting orbit which is an elliptical one; in this case we have the solution
\bes\label{solOeliptica}
\bal
\psi_e(x) &= \tanh(2rx),\label{solOelipA}\\
\chi_e(x) &= \sqrt{\frac{1\!-\!2r}{r}}\,\sech(2rx),\label{solOelipB}
\eal
\ees
with $r\in (0, 1/2]$. The parameter $r$ controls the thickness of the solutions and the height of $\chi_e(x)$, as can be seen in Fig.~\ref{fig01}, where we display the above solutions for $r=0.15$, $0.30$ and $0.45$. We can to show that for $r\to 1/2$ the orbit becomes a straight line, with the solution
\bes\label{solOreta}
\bal
\psi_s(x) &= \tanh(x),\label{solretaA}\\
\chi_s(x) &= 0.\label{solretapB}
\eal
\ees

\begin{figure}[t]
    \begin{center}
        \includegraphics[scale=0.6]{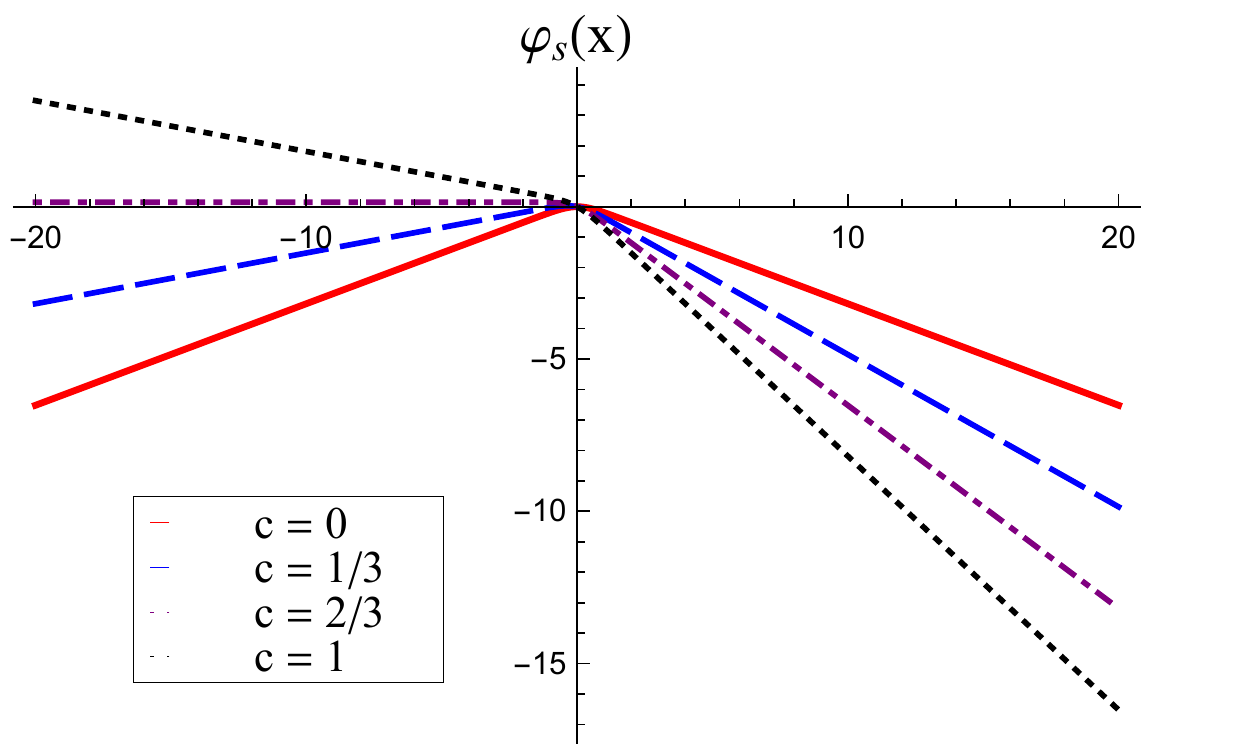}
        \includegraphics[scale=0.6]{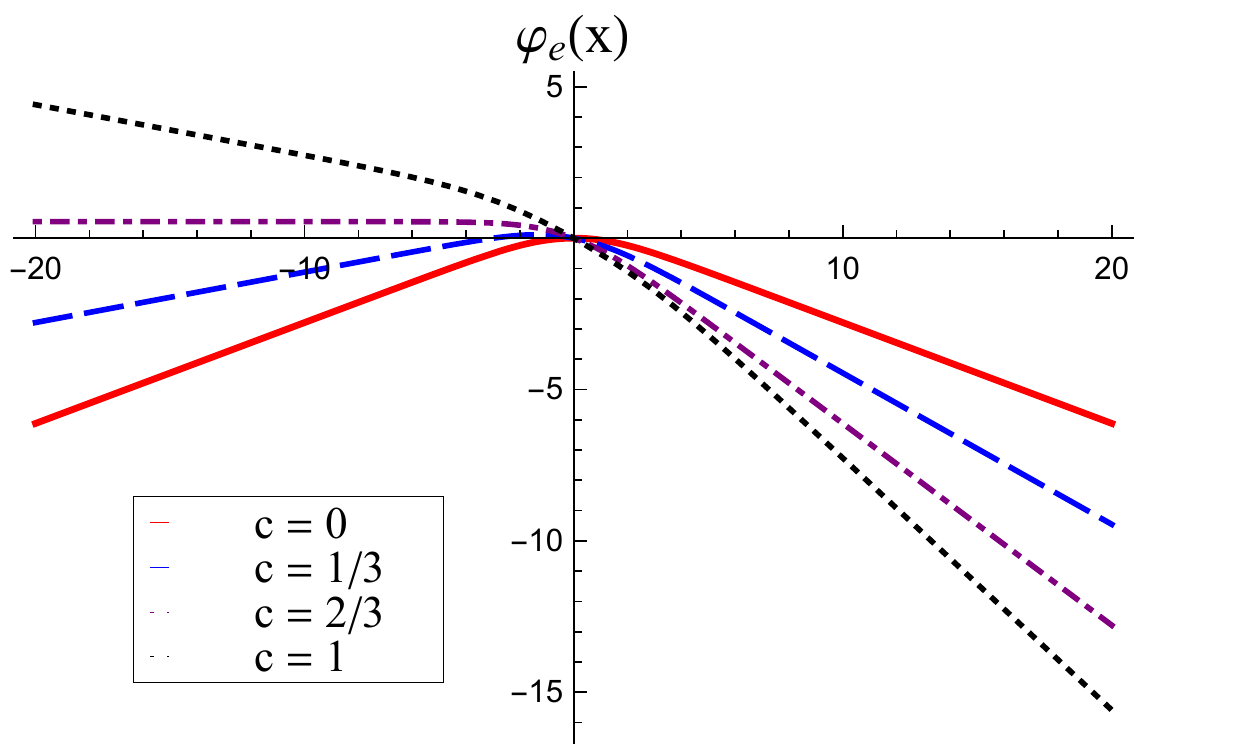}
    \end{center}
    \vspace{-0.5cm}
    \caption{\small{Dilaton field for straight orbit (top panel) and for elliptic orbit (bottom panel) with $\kappa=1$ and $r=1/4$.} \label{fig02}}
\end{figure}

By use of Eq. \eqref{foA} and \eqref{solD} we can obtain the dilaton field solution for the different possible orbits. For example, for the straight orbit we get
\be\label{dilatonreta}
\varphi_s(x) = -\frac{\kappa c}{2}x +\frac{\kappa}{3}\ln\left(\sech(x)\right) -\frac{\kappa}{12}\tanh^2(x).
\ee
On the other hand, for the elliptic orbit we have
\be\label{dilatonelipt}
\begin{aligned}
\!\!\varphi_e(x) \!=\! -\frac{\kappa c}{2}x \!+\!\frac{\kappa}{6r}\!\ln\!\left(\sech(2rx)\right)
\!+\!\frac{\kappa(1\!-\!3r)}{12r}\!\tanh^2\!(2rx).
\end{aligned}
\ee
Moreover, Fig. \ref{fig02} shows the behavior of the dilaton field solution obtained by previous equations and depicted for $\kappa=1$, $r=1/4$ (elliptical orbit) and $c=0,\,1/3,\,2/3,\,1$.

\begin{figure}[t]
    \begin{center}
        \includegraphics[scale=0.6]{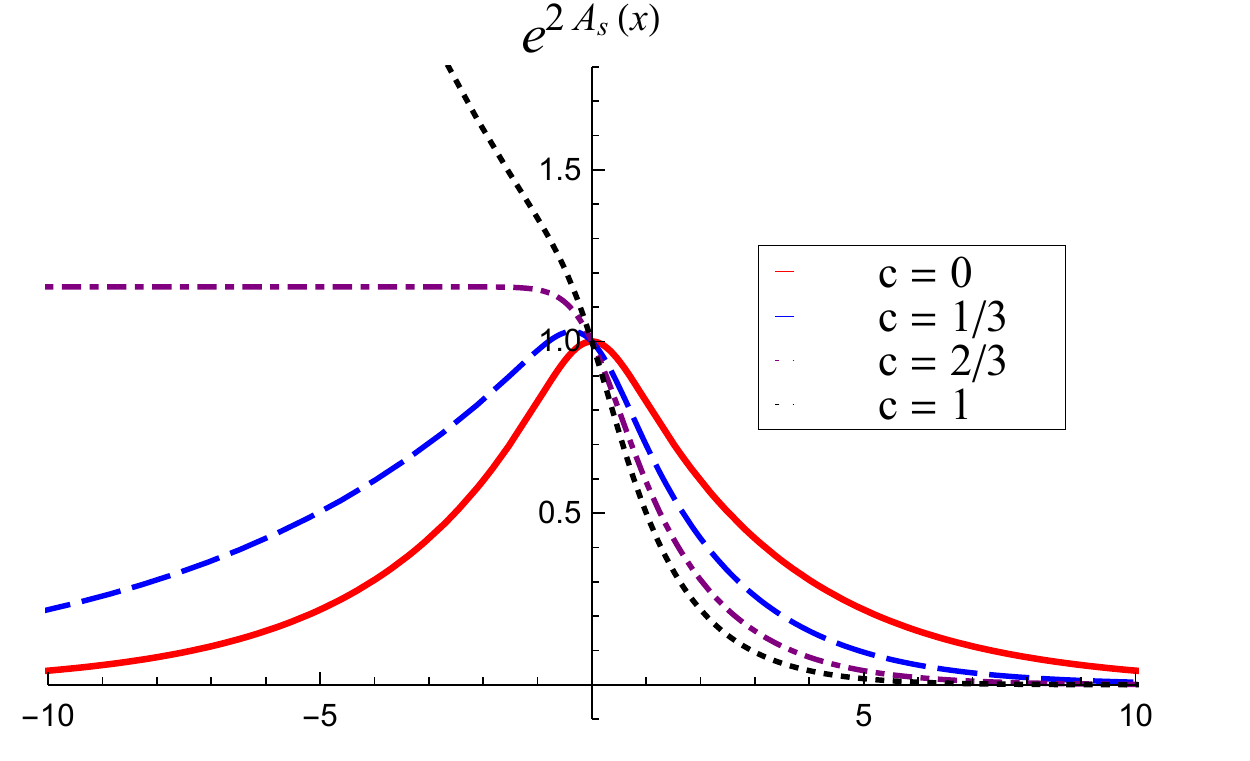}
        \includegraphics[scale=0.6]{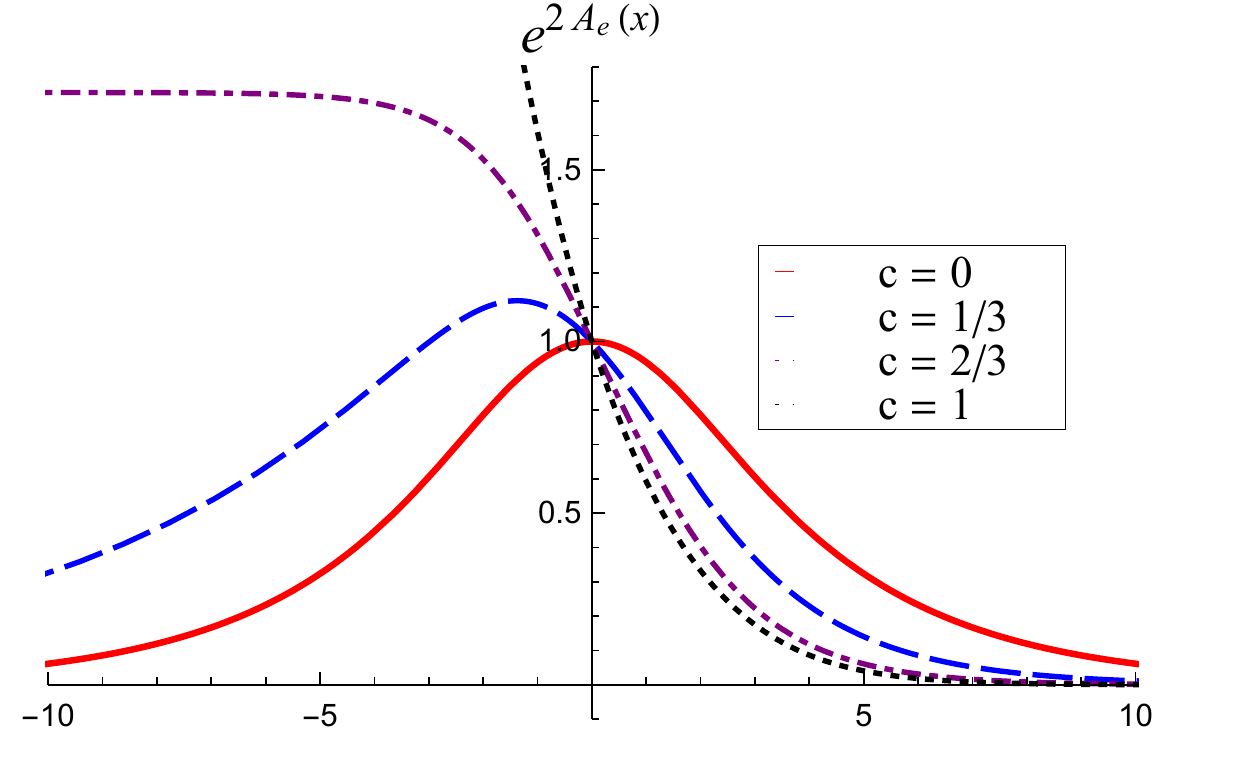}
    \end{center}
    \vspace{-0.5cm}
    \caption{\small{Warp factor for straight orbit (top panel) and for elliptic orbit (bottom panel) with $\kappa=1$ and $r=1/4$.}\label{fig03}}
\end{figure}

We also study the warp factor, and the two panels in Fig. \ref{fig03} show how it behaves for straight orbit (top panel) and for elliptic orbit (bottom panel) with $r=1/4$.

\begin{figure}[h]
    \begin{center}
        \includegraphics[scale=0.6]{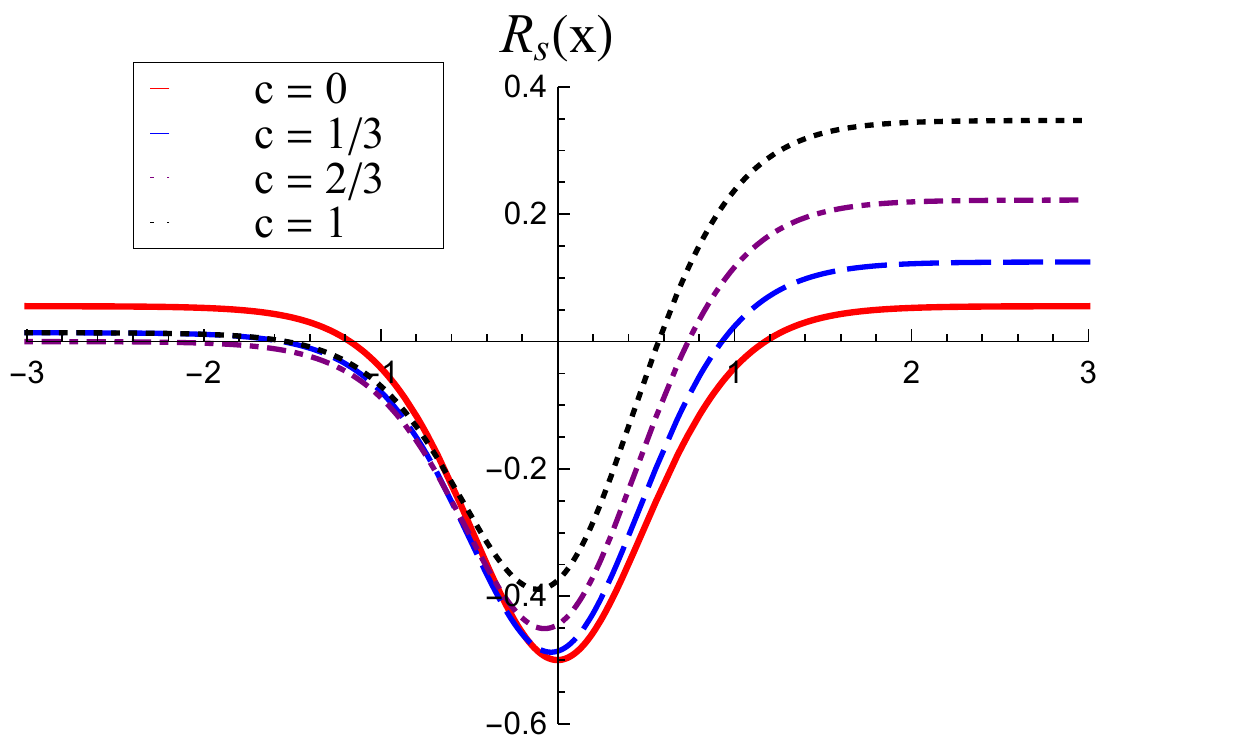}
        \includegraphics[scale=0.6]{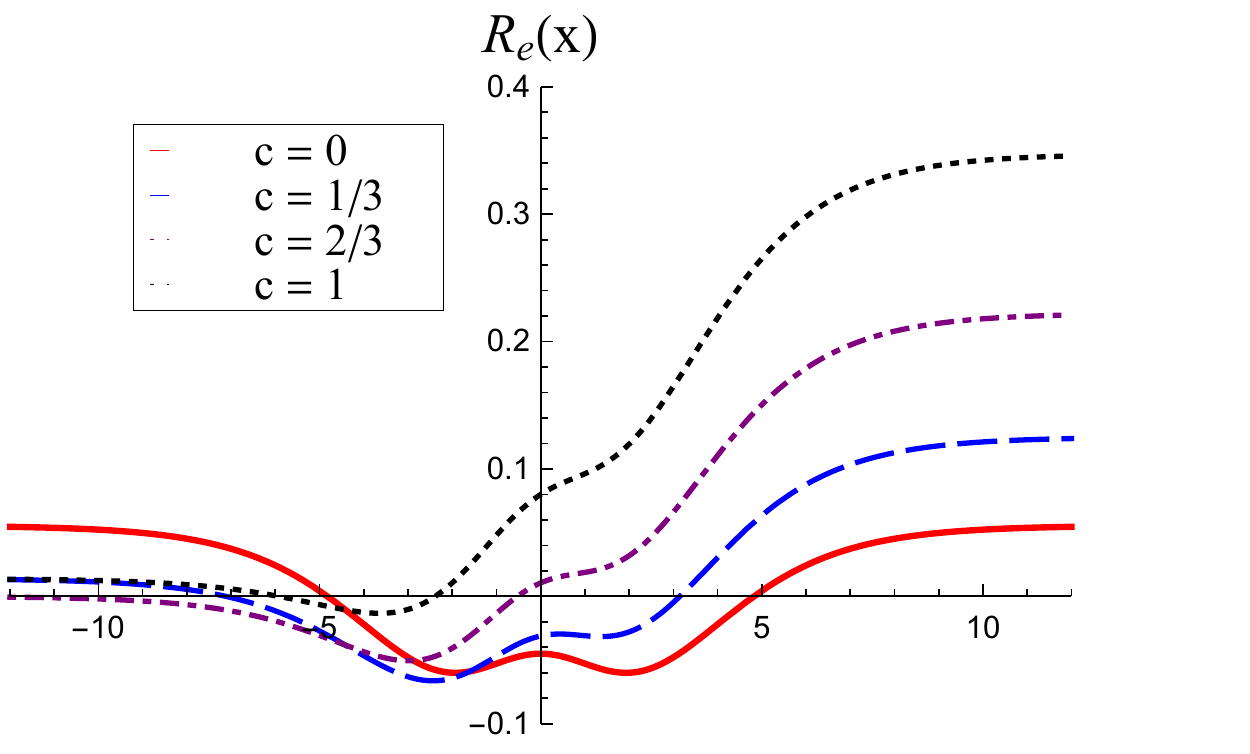}
    \end{center}
    \vspace{-0.5cm}
    \caption{\small{Ricci scalar for straight orbit (top panel) and for elliptic orbit (bottom panel) with $\kappa=1$ and $r=1/4$.}\label{fig04}}
\end{figure}

We can also calculate the Ricci scalar to verify how it behaves for the orbits obtained above. Using the Eq.~\eqref{Ricci} and the potential in the Eq.~\eqref{VBNRT}, the Ricci scalar can be obtained: for the straight orbit we get
\be
\!\!R_s(x) \!=\! -\frac{\kappa}{2}\,\sech^4(x) \!+\!\frac{\kappa^2}{8}\!\left(\!c \!+\!\tanh(x)\!-\!\frac13\!\tanh^3(x)\!\right)^2\!\!.
\ee
On the other hand, in the case of the elliptic orbit we obtain 
\ben
\!R_e(x) \!&=&\! -2\kappa r\left((1\!-\!2r)\sech^2(2rx)\!-\!(1\!-\!3r)\sech^4(2rx)\right)\nn
	&&+\frac{\kappa^2}{8}\bigg(c +\tanh(2rx)-\frac13\tanh^3(2rx)\\ 
	&&-(1 -2r)\,\sech^2(2rx)\tanh(2rx)\bigg)^2.\nonumber
\een
Fig. \ref{fig04} shows the Ricci scalar for the two orbits obtained here. We used $\kappa=1$, $c=0,1/3,2/3,1$ and for the elliptic orbits, $r=1/4$. For both straight and elliptic orbit we have that $R(x\!\to\!\pm \infty)\!=\!(\kappa^2/72)(2\pm3c)^2$, as calculated below Eq.~\eqref{VBNRT}. As we can see, the Ricci scalar is constant asymptotically, indicating that the two-dimensional space can be $M_2$ or $AdS_2$, see Ref. \cite{Cadoni:1999ja} for more details.

Now we turn attention to investigate the linear stability of the solutions. We begin by analyzing the stability of the solutions of the straight orbit given by Eq.~\eqref{solOreta}. In this case, the component $q(x)$ of the Eqs. \eqref{compstabSL} vanish, i.e., $q(x)=0$. Thus, the equations of stability \eqref{stabSL} become two independent equations. In this case, we can examine each perturbation separately, i.e.,
\bes\label{stabreta}
\bal
-e^{A}\left(e^{A}\eta_n'\right)'+e^{2A}p(x)\eta_n =\,& \omega_n^2\eta_n,\label{stabreta1}\\
-e^{A}\left(e^{A}\xi_m'\right)'+e^{2A}\bar{p}(x)\xi_m =\,& {\omega}_m^2\xi_m,\label{stabreta2}
\eal
\ees
where 
\bes
\bal
\!\!\!\!p(x)\! =&\, 4 \!-\!6\,\sech^2(x) \!+\!\frac{\kappa}{2}W_s(x)\tanh(x)\!+\!\frac{\kappa}{2}\sech^4(x)\nn
&+2\!\left(\!\frac{4\tanh(x)}{W_s(x)} \!+\!\frac{\,\sech^4(x)}{ W_s^3(x)}\!\right)\!\sech^4(x),\\
\!\!\!\!\bar{p}(x) \!=&\, 1-2\,\sech^2(x) +\frac{k}{4}W_s(x)\tanh(x),
\eal
\ees
and we defined $W_s(x)=c+\tanh(x)-(1/3)\tanh^3(x)$. We can use the Eq. \eqref{stabreta1} to obtain the zero mode as
\be\label{zeroreta1}
\eta_0(x) = \Nc\,\frac{\sech^2(x)}{W_s(x)},
\ee
where $\Nc$ is a normalization constant that can be obtained of the Eq. \eqref{norma}. It is possible to verify that the zero mode is only normalizable for $c>2/3$. This integration can be done numerically, for example, when $\kappa\!=\!c\!=\!1$ we have that $\Nc\approx 0.668$. Using the supersymmetric partner equation in \eqref{partnersusy}, it is possible to show that there is no normalizable eigenstate, regardless of the value of $c$. Thus, the stability equation \eqref{stabreta1} will only have the eigenvalue $\omega_0=0$ with eigenstate given by \eqref{zeroreta1}.

\begin{figure}[b]
    \begin{center}
        \includegraphics[scale=0.6]{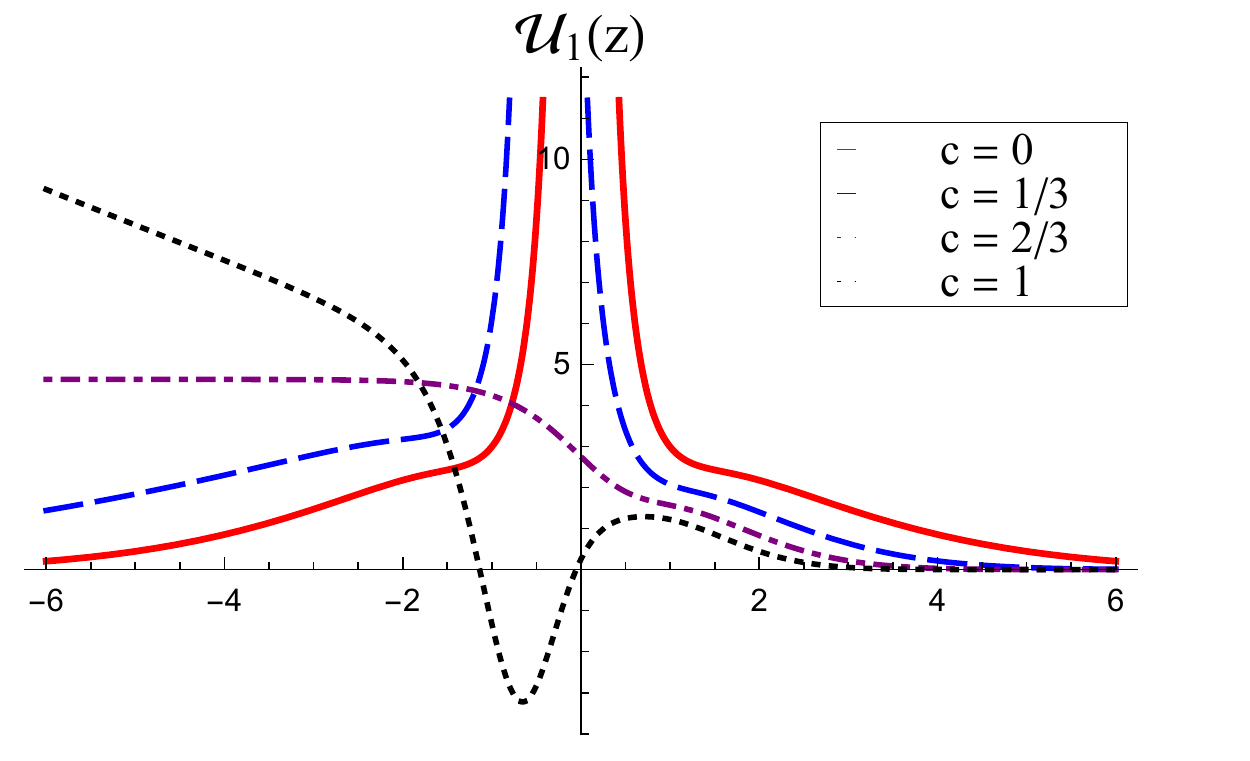}
            \hspace{0.2 cm}
        \includegraphics[scale=0.6]{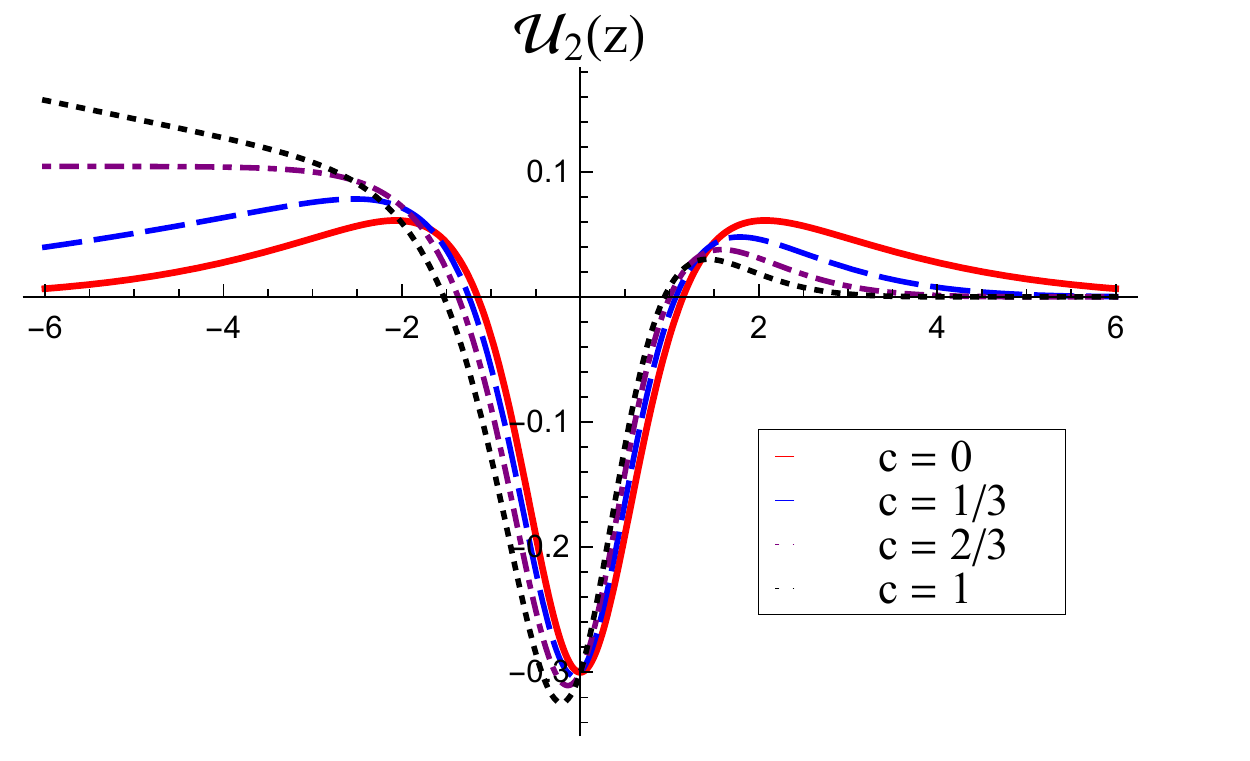}
    \end{center}
    \vspace{-0.5cm}
    \caption{\small{Plot of the stability potentials ${\cal U}_1(z)$ (top panel) and ${\cal U}_2(z)$ (bottom panel) for
    the straight orbit using $\kappa=1$.}\label{fig05}}
\end{figure}

Let us now analyze the Eq. \eqref{stabreta2}. In this case, the mode-zero is given by
\be\label{zeroreta2}
\xi_0(x) = \tilde{\Nc}\,\sech(x),
\ee
where $\tilde{\Nc}$ is another normalization constant. In this case, we find that the zero mode is always normalizable, regardless of the value of parameter $c$. For instance, taking $\kappa=1$ and $c=0$, $1/3$, $2/3$ and $1$, we obtain $\tilde{\Nc}\approx 0.683$, $0.682$, $0.679$ and $0.673$, respectively. We also verify that the zero mode is the only eigenstate present in this case.

We can change variables from $x$ to $z$, as $dz=e^{-A}dx$, so that the Eqs. \eqref{stabreta} became Schr\"odinger-like equations of the form
\bes
\bal
-\frac{d^2\eta_n}{dz^2}+{\cal U}_1(z)\eta_n =\,& \omega_n^2\eta_n,\\
-\frac{d^2\xi_m}{dz^2}+{\cal U}_2(z)\xi_m =\,& {\omega}_m^2\xi_m,
\eal
\ees
where the potentials are defined as ${\cal U}_1(z)=e^{2A(z)}p(z)$ and ${\cal U}_2(z)=e^{2A(z)}\bar{p}(z)$, and $A(z)=\varphi(z)/2$ is obtained by Eq. \eqref{dilatonreta}. In Fig. \ref{fig05} we show the behavior of the potentials  ${\cal U}_1(z)$ and ${\cal U}_2(z)$ for $\kappa=1$ and $c$ as in Fig.~\ref{fig02}. This result is numerical because it is not possible to change variable from $x$ to $z$ analytically. It is possible to show that the potential ${\cal U}_1(z)$ supports zero mode if $2/3<c<2(12+\kappa)/(3\kappa)$. This result is in accordance with the investigation presented in Ref.~\cite{Zhong:2021gxs}. For the second field there is the potential ${\cal U}_2(z)$ that comes as result of the perturbation around the corresponding solution. In this case, it is possible to emerge zero mode if $2/3<c<2(6+\kappa)/(3\kappa)$, as it can be seen in the bottom panel of Fig. \ref{fig05}.

To study the stability for the elliptical orbit, we first notice that $q(x)\neq0$ and the system of Eqs. \eqref{stabSL} no longer decouple. In this case, we must deal with a matrix representation of the equations. Since the expressions of $p(x)$, ${\bar p}(x)$ and $q(x)$ in Eq. \eqref{stabSL} are long and awkward, we first define the quantities  $T(x)=\tanh(2rx)$, $S(x)=\sech(2rx)$ and
\be
\begin{aligned}
W_e(x) =&\, c +\tanh(2rx) -\frac13\tanh^3(2rx)\\
	&-(1-2r)\tanh(2rx)\,\sech^2(2rx)\\
=&\, c +T(x) -\frac13T^3(x)\!-\!(1\!-\!2r)T(x)\,S^2(x),
\end{aligned}
\ee
and now the components of Eq.~\eqref{compstabSL} can be worked out to be written in the following form
\bes
\bal
p(x) &= 4\!-4\left(1+2r^2\right)\!S^2(x)+\kappa r^2S^4(x)\nn
\!\!&\!+\frac{\kappa}{2}W_e(x)T(x)+ 32r^3S^4(x)\Bigg(\frac{2T(x)}{W_e(x)}\nn
\!\!&\!+\frac{\left(1\!-\!2r\!-\!(1\!-\!3r)S^2(x)\right)\!S^2(x)}{ W_e^2(x)}\Bigg)\!,\\
\!\!\bar{p}(x) &= 4r^2 +4r(1-4r)S^2 +\frac{\kappa r}{2}W_e(x)T(x)\nn
\!\!&+\!\kappa r(1-2r)S^2(x)T(x)\!\left(\!T(x) \!+\!\frac{32r\!\left(1 \!-\!2S^2(x)\right)}{\kappa W_e(x)}\right.\nn
\!\!&\left.+\frac{32r\!\left(1\!-2r \!-\!(1\!-\!3r)S^2(x)\!\right)\!S^2(x)T(x)}{\kappa W_e^2(x)}\!\right)\!,\\
\!\!q(x) &= \sqrt{r(1\!-\!2r)}\Bigg(\!\!\left(4(1\!+\!2r)\!+\!\kappa rS^2(x)\right)\!S(x)T(x)\nn
\!\!&+\!\frac{\kappa}{2}W_e(x)S^2(x)\!-\!2\kappa rS^3(x)\!\bigg(\!T(x)\!+\!\frac{8r\!\left(3\!-\!4S^2(x)\right)}{\kappa W_e(x)}\nn
\!\!&+\frac{16r\!\left(1\!-\!2r\!-\!(1\!-\!3r)S^2(x)\right)S^2(x)T(x)}{\kappa W_e(x)}\bigg)\Bigg).
\eal
\ees

In this case, the the zero mode can be obtained by Eq. \eqref{zeroSL} in the form
\be
\!\!\!\Upsilon^{(0)}(x)\! =\! \frac{\Nc\sech(2rx)}{W_e(x)}\!
\begin{pmatrix}
    \sech(2rx)\vspace{0.2cm}\\
    -\sqrt{\frac{1-2r}{r}}\,\tanh(2rx) \\
\end{pmatrix}\!,
\ee
where $\Nc$ is the normalization constant determined by Eq. \eqref{norma} and in order to have normalized states we need to assume that $c>2/3$.

\subsection{Model B}\label{modB}

Let us now consider another model where we obtain kink-like solutions. For this, we assume that
\ben\label{WmodB}
W(\psi,\chi) = c +\gamma_1\!\left(\!\psi-\frac13\psi^3\!\right) +\gamma_2\!\left(\!\chi-\frac13\chi^3\!\right)\!,
\een
where, in addition to the real constant $c$, we also introduce two new real parameters $\gamma_1$ and $\gamma_2$ that influence the thickness of the solutions. It is interesting to note that although the Eq. \eqref{WmodB} does not present interaction between fields explicitly, we still have a system whose potential $V(\psi,\chi)$ is coupled. This can be easily verified by Eq. \eqref{potW} whose term $W^2$ provides interaction between the two fields. This possibility was also considered in Refs. \cite{Bazeia:2006ef,Ahmed:2012nh,deSouzaDutra:2014ddw} in the braneworld context in five-dimensional spacetime with an extra spatial dimension of infinite extent. In the present case, the potential becomes
\be\label{potMB}
\begin{aligned}
\!V(\psi,\chi)\! &= \frac{\gamma_1^2}{2}\left(1-\psi^2\right)^2 +\frac{\gamma_2^2}{2}\left(1 -\chi^2\right)^2\\
&-\frac{\kappa}{8}\left(\!c+\gamma_1\!\left(\!\psi-\frac13\psi^3\!\right)+\gamma_2\!\left(\!\chi-\frac13\chi^3\!\right)\right)^2\!.
\end{aligned}
\ee
The asymptotic values of solutions of model \eqref{WmodB} can also be obtained by algebraic equations in the form $W_\psi=0$ and $W_\chi=0$. In this case, we have $v_{\pm}=(1,\pm1)$ and $\bar{v}_{\pm}=(-1,\pm1)$. Thus, $W(v_{\pm})\!\equiv\! W_{\pm} \!=\!c+2\gamma_1/3\pm2\gamma_2/3$ and $W(\bar{v}_{\pm})\equiv \bar{W}_{\pm}=c-2\gamma_1/3\pm2\gamma_2/3$. Using the asymptotic values of solutions in potential, we obtain $V(v_{\pm})=- \kappa  W_{\pm}^2/8$ and $V(\bar{v}_{\pm})=-\kappa  \bar{W}_{\pm}^2/8$.

\begin{figure}[t]
    \begin{center}
        \includegraphics[scale=0.6]{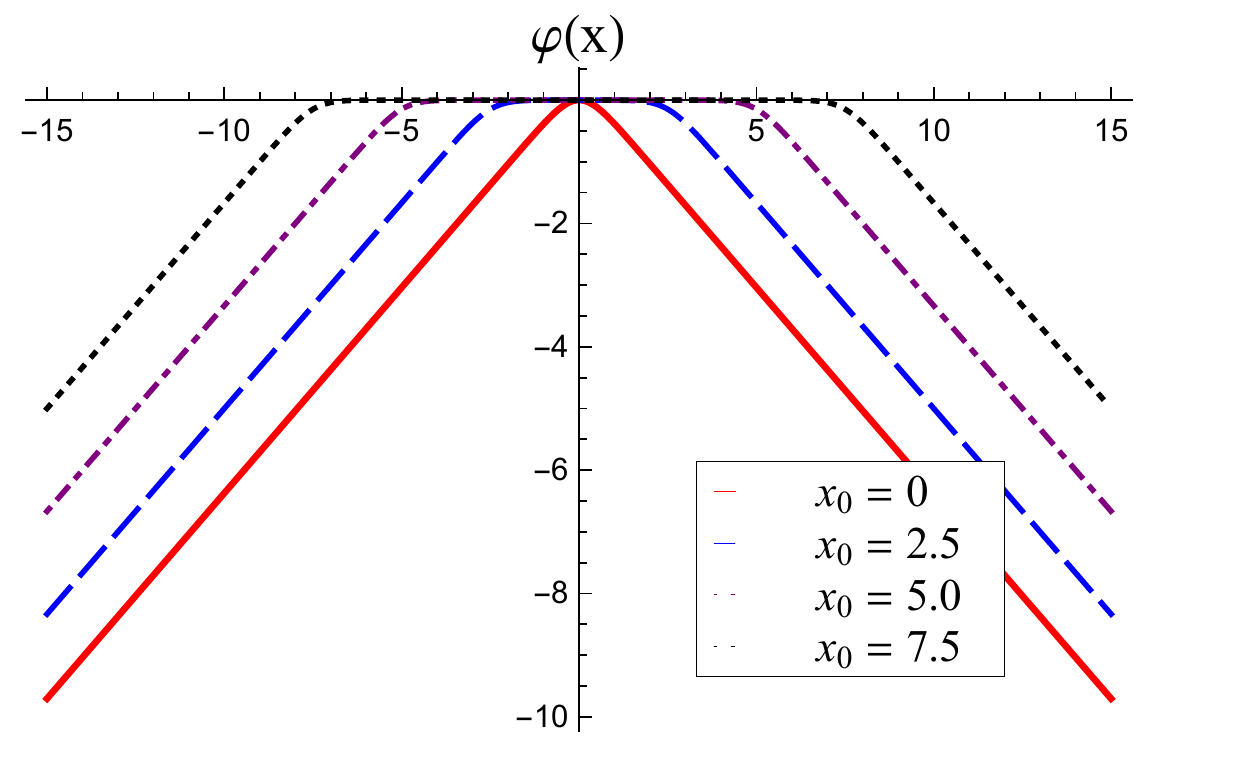}
        \includegraphics[scale=0.6]{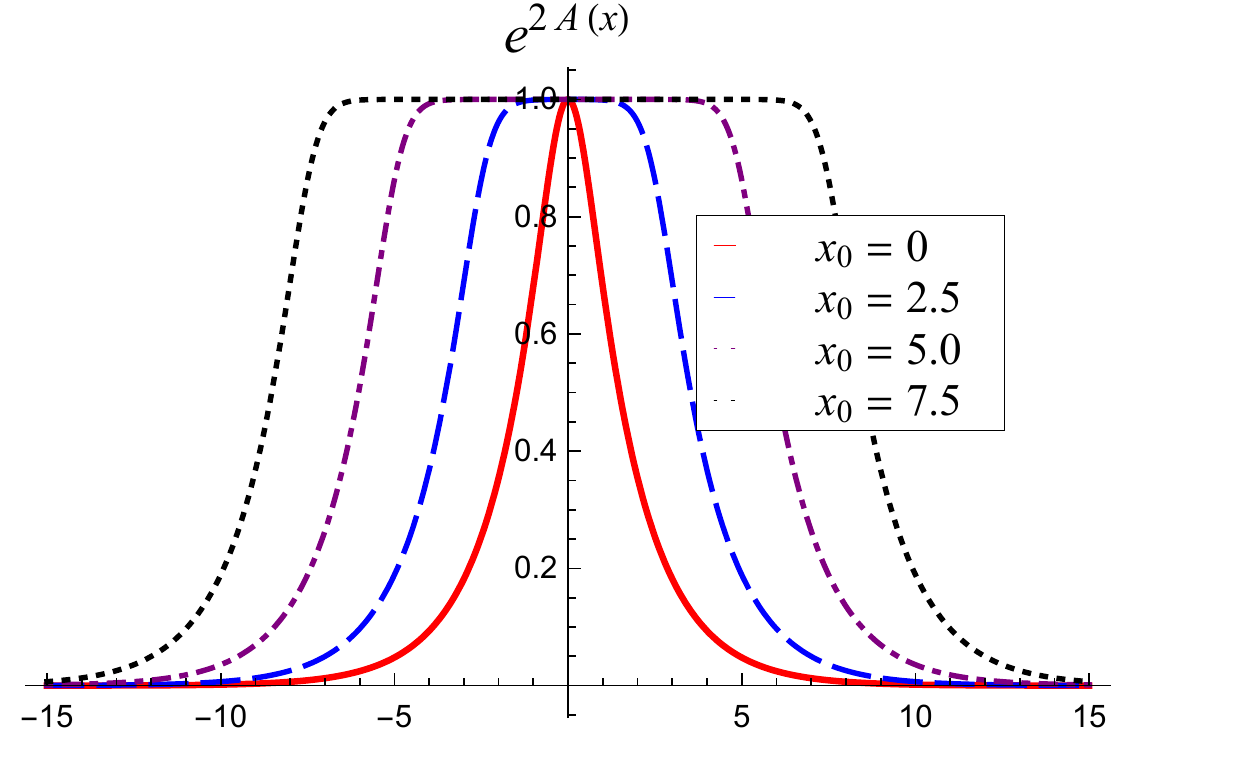}
    \end{center}
    \vspace{-0.5cm}
    \caption{\small{Dilaton field and warp factor plotted for $c=0$, $\kappa=1$, $\gamma_1=\gamma_2=1$ and $\tilde{x}_0=-x_0$.}\label{fig06}}
\end{figure}

\begin{figure}[t]
    \begin{center}
        \includegraphics[scale=0.6]{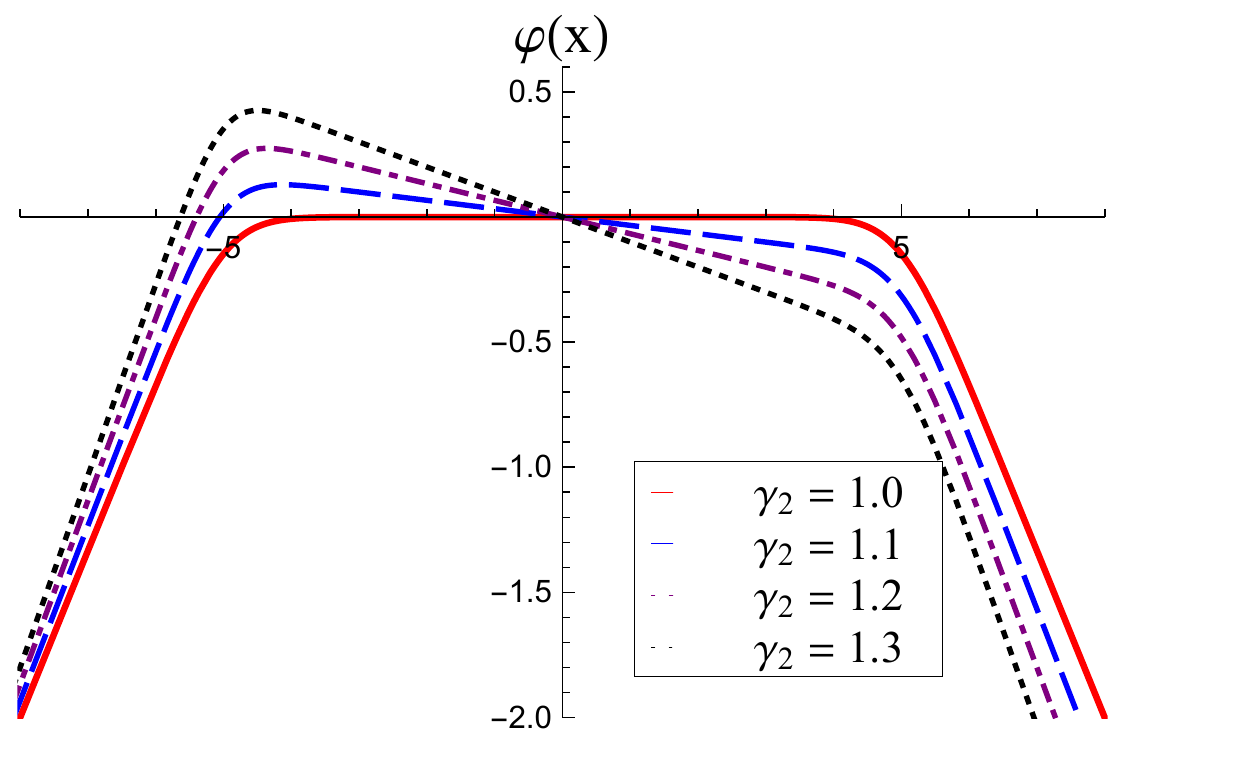}
        \includegraphics[scale=0.6]{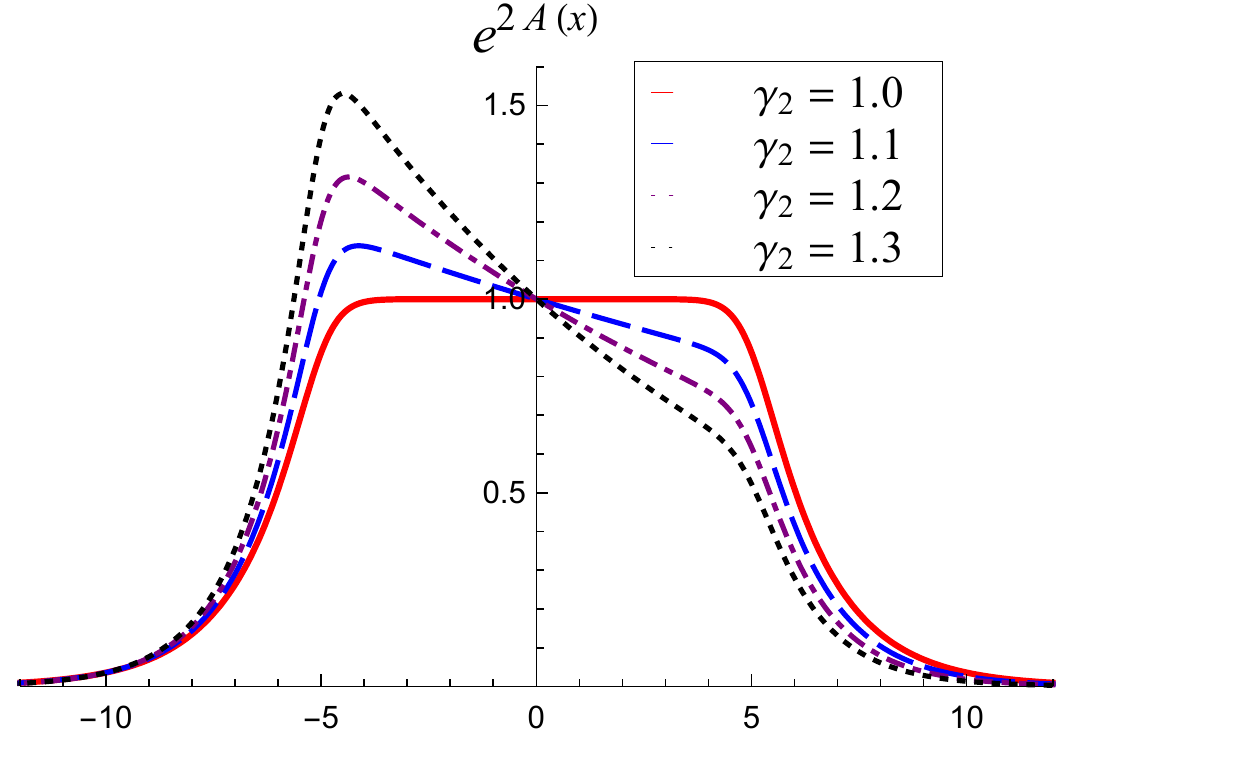}
    \end{center}
    \vspace{-0.5cm}
    \caption{\small{Dilaton field and warp factor plotted for $c=0$, $\kappa=\gamma_1=1$ and $x_0=-\tilde{x}_0=5$.}\label{fig07}}
\end{figure}

Let us now obtain the solutions for this model. We can write the first-order equations \eqref{fo} as
\bes
\bal
\psi^\prime &= \gamma_1\left(1 -\psi^2\right),\\
\chi^\prime &= \gamma_2\left(1 -\chi^2\right).
\eal
\ees
The above first-order equations allow kink-like solutions in the form,
\bes
\bal
\psi(x) &= \tanh\big(\gamma_1(x-x_0)\big),\\
\chi(x) &= \tanh\big(\gamma_2(x-\tilde{x}_0)\big),
\eal
\ees
where $x_0$ and $\tilde{x}_0$ are real constants that defines the center of the solutions, while the parameters $\gamma_1$ and $\gamma_2$ control their thickness. We can also find the dilaton field using Eqs.~\eqref{foA} and \eqref{solD} to write
\ben\label{WFMB}
\begin{aligned}
\!\!\!\varphi(x) \!=& -\frac{\kappa c}{2}x \!+\!\frac{\kappa}{3}\ln\!\left(\!\frac{\sech(\gamma _1(x\!-\!x_0))\,\sech(\gamma _2(x\!-\!\tilde{x}_0))}{\sech(\gamma_1x_0)\,\sech(\gamma_2\tilde{x}_0)}\!\right)\\
\!\!\!&-\frac{\kappa}{12}\Big(\tanh^2(\gamma_1(x-x_0)) -\tanh^2(\gamma_1x_0)\\
\!\!\!&+\tanh^2(\gamma_2(x-\tilde{x}_0)) -\tanh^2(\gamma_2\tilde{x}_0)\Big).
\end{aligned}
\een
As we know it is possible to calculate the warp factor of this model from the above equation as $e^{2A(x)}$. Let us now turn our attention to how the parameters $c$, $\gamma_1$, $\gamma_2$, ${x}_0$ and $\tilde{x}_0$ modify the dilaton field and warp factor. Firstly, we consider $c=0$, $\gamma_1=\gamma_2$ and $\tilde{x}_0=-x_0$; in this case the parameter $x_0$ enlarges the center of the dilaton and warp factor. This behavior can be seen in Fig. \ref{fig06}, where we display $\varphi(x)$ and $e^{2A(x)}$, for $c=0$, $\kappa=\gamma_1=\gamma_2=1$ and $x_0=0$, $2.5$, $5$ and $7.5$. In the previous model, the parameter $c$ was responsible for an asymmetry in the dilaton field and warp factor, however in the present model we can generate an asymmetry from the parameters $\gamma_1$ and $\gamma_2$. To see this, we take in Fig. \ref{fig07}, $c=0$ and $\kappa=\gamma_1=1$, $x_0=-\tilde{x}_0=5$ with $\gamma_2=1$, $1.1$, $1.2$ and $1.3$. As we see, it is possible to obtain asymmetric quantities with the parameter $c=0$. This is interesting because the model allows the warp factor to be always localized. Nevertheless, it is also possible to obtain asymmetric quantities through the parameter $c$. However, depending on the value of $c$ it may lead to delocalized warp factor. In addition, the parameter $c$ works differently from $\gamma_1$ and $\gamma_2$ to produce asymmetries, therefore it is not possible to cancel each other's effects.

\begin{figure}[t]
    \begin{center}
        \includegraphics[scale=0.6]{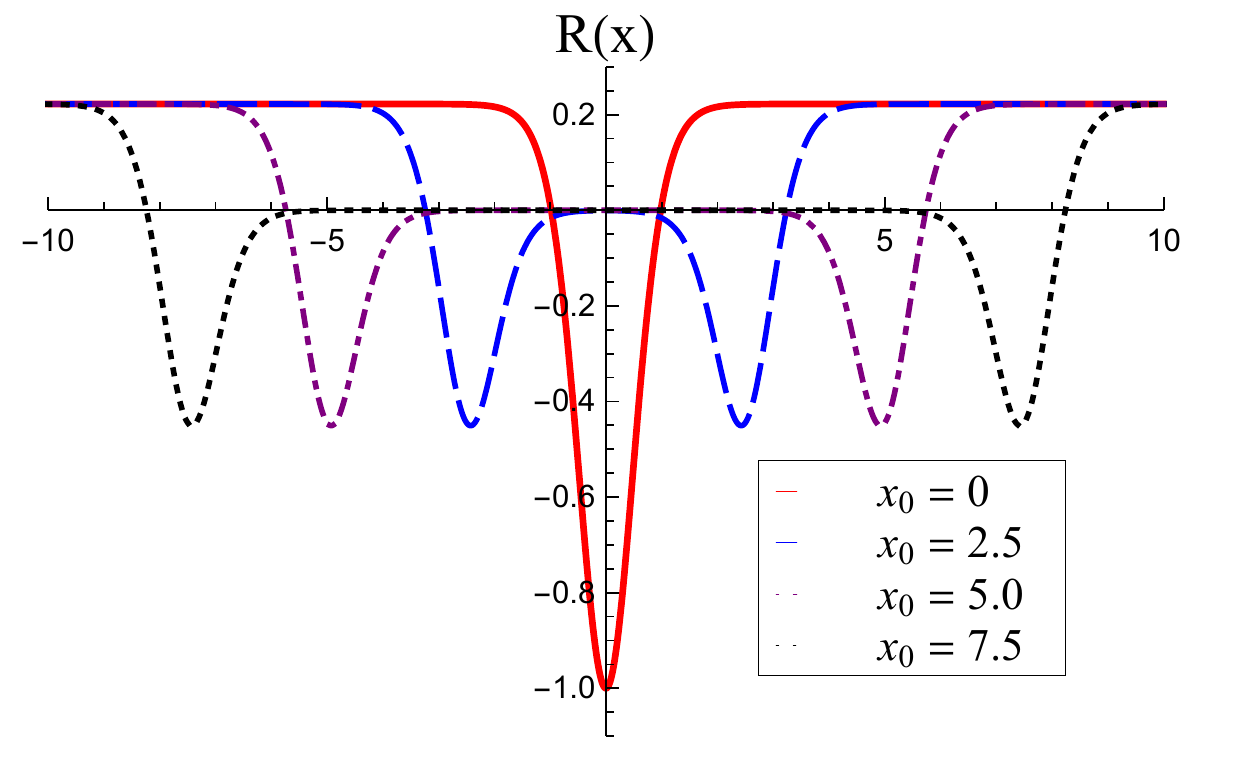}
        \includegraphics[scale=0.6]{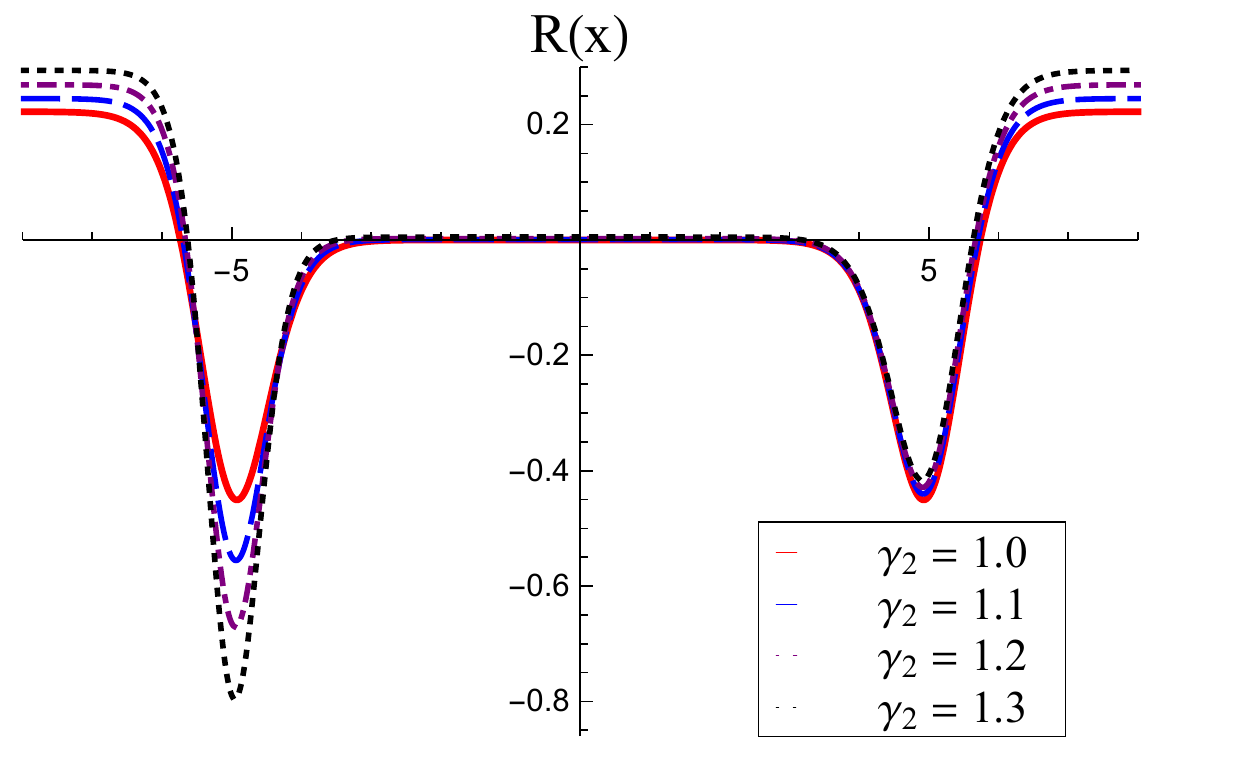}
    \end{center}
    \vspace{-0.5cm}
    \caption{\small{Top panel shows the Ricci scalar depicted for $c=0$, $\kappa=\gamma_1=\gamma_2=1$ and $\tilde{x}_0=-x_0$. Bottom panel shows the Ricci scalar depicted for $c=0$, $\kappa=\gamma_1=1$ and $x_0=-\tilde{x}_0=5$.}\label{fig08}}
\end{figure}

To better understand the asymptotic behavior of gravity we can calculate the Ricci scalar. In the present case we have,
\be\label{RSMB}
\begin{aligned}
R(x) &= -\frac{\kappa\gamma_1^2}{2}\sech^4\left(\gamma_1(x\!-\!x_0)\right) -\frac{\kappa\gamma_2^2}{2}\sech^4\left(\gamma_2(x\!-\!\tilde{x}_0)\right)\\
&+\!\frac{\kappa^2}{8}\!\bigg(\!c \!+\!\gamma_1\!\tanh\!\left(\gamma_1(x\!-\!x_0)\right) \!-\!\frac{\gamma_1}3\!\tanh^3\!\left(\gamma_1(x\!-\!x_0)\right)\\
&+\!\gamma_2\!\tanh\!\left(\gamma_2(x\!-\!\tilde{x}_0)\right) 
\!-\!\frac{\gamma_2}3\!\tanh^3\!\left(\gamma_2(x\!-\!\tilde{x}_0)\right)\!\!\bigg)^2\!.
\end{aligned}
\ee
It is possible to show that,
\ben\nonumber
\lim_{x\to\pm\infty} R(x)\to \frac{\kappa ^2}{8}  \left(\frac23|\gamma _1| +\frac23|\gamma _2| \pm c\right)^2\,.
\een
This result show that the Ricci scalar tends to a non negative constant asymptotically, where the parameter $c$ also generates an asymmetry in this case. In Fig.~\ref{fig08} we depicted the Ricci scalar for some values of the parameters. In the top panel we use $c=0$, $\kappa=\gamma_1=\gamma_2=1$, $\tilde{x}_0=-x_0$ with $x_0$ as in Fig.~\ref{fig06}. In the bottom panel we use $c=0$, $\kappa=\gamma_1=1$, $x_0=-\tilde{x}_0=5$ and $\gamma_2$ as in Fig.~\ref{fig07}.

Again, the study of stability is implemented via the matrix Eq. \eqref{stabSL}, with the components \eqref{compstabSL}, which in this case can be written as,
\bes
\bal
\!\!\!p(x) \!&=\! 4\gamma_1^2\!-\!6\gamma_1^2\!+\!\frac{\kappa\gamma_1}{2}W(x)T_1(x)+\frac{\kappa\gamma_1^2}{4}S_1^4(x)\nn
\!\!\!&+\!2\gamma_1^2S_1^4(x)\bigg(\!\frac{4\gamma_1T_1(x)}{W(x)} \!+\!\frac{\gamma_1^2S_1^4(x) \!+\!\gamma_2^2S_2^4(x)}{W^2(x)}\!\bigg),\\
\!\!\!\!\!\!\bar{p}(x) \!&=\! 4\gamma_2^2\!-\!6\gamma_2^2\!+\!\frac{\kappa\gamma_2}{2}W(x)T_2(x) \!+\!\frac{\kappa\gamma_2^2}{4}\!S_2^4(x)\nn
\!\!\!&+\!2\gamma_2^2S_2^4(x)\left(\!\frac{4\gamma_2T_2(x)}{W(x)} \!+\!\frac{\gamma_1^2S_1^4(x) \!+\!\gamma_2^2S_2^4(x)}{W^2(x)}\!\right),\\
\!\!\!q(x) \!&=\! \frac{\kappa\gamma_1\gamma_2}{4}\Bigg(\!1\!+\!\frac{16\left(\gamma_1T_1(x)\!+\!\gamma_2T_2(x)\right)}{\kappa W(x)}\nn
\!\!\!&+\frac{8\left(\gamma_1^2S_1^4(x)\!+\!\gamma_2^2S_2^4(x)\right)}{\kappa W^2(x)}\!\Bigg)S_1(x)S_2(x),
\eal
\ees
where $T_1(x)\!=\!\tanh\big(\gamma_1(x\!-\!x_0)\big)$, $T_2(x)\!=\!\tanh\big(\gamma_2(x\!-\!\tilde{x}_0)\big)$, $S_1(x)=\sech\big(\gamma_1(x\!-\!x_0)\big)$, $S_2(x)=\sech(\gamma_2\big(x\!-\!\tilde{x}_0)\big)$ and 
\be
\begin{aligned}
\!\!\!\!\!W(x) \!=\,& c \!+\!\gamma_1\!\left(\!\tanh\!\big(\gamma_1(x\!-\!x_0)\big) \!-\!\frac13\tanh^3\!\big(\gamma_1(x\!-\!x_0)\big)\!\right)\\
\!\!\!\!\!&\!+\!\gamma_2\!\left(\!\tanh\!\big(\gamma_2(x\!-\!\tilde{x}_0)\big) \!-\!\frac13\tanh^3\!\big(\gamma_2(x\!-\!\tilde{x}_0)\big)\!\right)\\
=\,& c \!+\!\gamma_1\!\left(\!T_1(x) \!-\!\frac13T_1^3(x)\!\right)\!+\!\gamma_2\!\left(\!T_2(x) \!-\!\frac13T_2^3(x)\!\right).
\end{aligned}
\ee

Again, we can calculate the zero mode in the following form
\be
\Upsilon^{(0)}(x) = \frac{\Nc}{W(x)}
\begin{pmatrix}
    \gamma_1\sech^2\big(\gamma_1(x\!-\!x_0)\big)\vspace{0.2cm}\\
    \gamma_2\sech^2\big(\gamma_2(x\!-\!\tilde{x}_0)\big)\\
\end{pmatrix},
\ee
where the normalization constant $\Nc$ can be obtained using Eq.~\eqref{norma}.

\section{Comments and conclusions}\label{coments}

In this work, we studied two-dimensional Jackiw-Teitelboim gravity, where the Lagrange density of matter displays coupled scalar fields. We investigated models that appeared before in the study of five-dimensions braneworld and considered two specific situations where it is possible to obtain topological solutions to the matter fields analytically. In each case, we obtained the solution of dilaton field and analyzed the linear stability. We verified that, in general, the equations of stability for the matter fields are coupled, being obtained in the matrix form.

In the first model, we investigated the analogous situation of the so-called Bloch brane that was studied in \cite{Bazeia:2004dh}. We verified that it is possible to obtain a set of solutions for the matter fields that present topological behavior and connect the asymptotic values obtained by algebraic equations $W_\psi=0$ and $W_\chi=0$, however, the result is generally non-analytical. Nevertheless, for an adequate choice of parameters, it is possible to obtain two classes of analytical solutions that described straight and elliptic orbits. For these two specific situations, we found the dilaton field solution and show that the Ricci scalar presents the usual behavior, connecting two asymptotically $AdS_2$ or $M_2$ spaces. Although the general solutions only allow us to reconstruct the stability potential in the matrix form, we found that for the specific case of the straight orbit the equations of perturbations decoupled and we could analyze each perturbation of matter fields separately.

In the second model studied in this paper, we used an auxiliary function that generates kink solutions with different thicknesses. We showed that, even though the coupling in the fields is not present in the auxiliary function, the scalar potential is still coupled. We also obtained the solution of the dilaton field and verified the behavior of the Ricci scalar. Moreover, we studied stability within is a matrix representation since the equations of stability do not decouple. However, we found the zero mode analytically.

In addition to the study presented here, we think it is also of current interest to address situations where the JT gravity is generalized by introducing new scalars built from the Ricci scalar, such as in $\,F(R)$-gravity \cite{Rippl:1995bg, Hwang:2001pu, DeFelice:2010aj, Sotiriou:2008rp}, and models with gauge and other fields. Another direction of current interest concerns the study of 2D Einstein-Maxwell-Dilaton gravity and connections with $AdS_2$ holography; see, e.g., Ref. \cite{Cvetic:2016eiv} and references therein for further details on this subject. Furthermore, modifications that aim to encompass exotic properties such as dark matter and dark energy were obtained through the inclusion of matter fields with unusual dynamics, called $K$-fields \cite{Armendariz-Picon:1999hyi, Armendariz-Picon:2000nqq}. We are also studying dilaton gravity \cite{Grumiller:2002nm} with the inclusion of fields that engenders compact behavior. We believe that new studies along the above lines may add other effects and give rise to new research perspectives for $2D$ gravity. These and other related issues are now under consideration, and we hope to report on them in the near future.

\begin{acknowledgments}

DB and RM would like to thank CNPq (Brazil), grants No. 303469/2019-6 (DB) and No. 310994/2021-7 (RM). They also thank Paraiba State Research Foundation, FAPESQ-PB, grants No. 0003/2019 (RM) and No. 0015/2019 (DB), for partial financial support.

\end{acknowledgments}


\end{document}